\documentstyle[amssymb,epsfig]{mn2e}

\title[Self-enrichment of Galactic halo globular clusters]
{Self-enrichment of Galactic halo globular clusters: stimulated star 
formation and consequences for the halo metallicity distribution}
\author[G.~Parmentier]{Genevi\`eve Parmentier
\thanks{E-mail: parm@astro.unibas.ch } \\
Institute of Astrophysics and Geophysics, University of Li\`ege, 
4000 Li\`ege, Belgium \\}
\begin{document}

\date{Accepted .... Received ... ; in original form ...}

\pagerange{\pageref{firstpage}--\pageref{lastpage}} \pubyear{2002}

\maketitle

\label{firstpage}

\begin{abstract}
We explore the self-enrichment hypothesis for globular cluster formation 
with respect to the star formation aspect.  
Following this scenario, the massive stars of a first stellar generation 
chemically enrich the globular progenitor cloud up to Galactic halo 
metallicities and sweep it into an expanding spherical 
shell of gas.  This paper investigates the ability of
this swept proto-globular cloud to become gravitationally unstable and, therefore, 
to seed the formation of second generation stars which may later on form 
a globular cluster.  We use a simple model based on a linear perturbation
theory for transverse motions in a shell of gas to demonstrate that the
pressures by which the progenitor clouds are bound and the 
supernova numbers required to achieve Galactic halo metallicities 
support the successful development of the shell transverse collapse.
Interestingly, the two parameters controling the metallicity achieved 
through self-enrichment, namely the number of supernovae and the external
pressure, also rule the surface density of the shell and thus its ability to 
undergo a transverse collapse.  Such a supernova-induced origin for the globular 
cluster stars opens therefore the way to the understanding of 
the halo metallicity distributions.  This model is also able to explain
the lower limit of the halo globular cluster metallicity.
\end{abstract}

\begin{keywords}
globular clusters: general -- Galaxy: halo -- supernova remnants -- 
stars: formation.
\end{keywords}

\section{Introduction}
Globular clusters (GC) are dense, massive and round-shaped
groups of stars present in the vast majority of galaxies.
In our Galaxy, the halo GCs were among the very first bound structures to 
form and their study provides therefore valuable information 
about the early Galactic evolution.  Their formation is an exciting but 
yet unsolved problem.  For instance, it is still an open question whether 
Galactic halo GCs formed out of 
gas already chemically enriched ({\it pre}-enrichment models, e.g. 
Harris \& Pudritz 1994) or whether they produced their 
own heavy elements through an earlier generation of stars within the 
GC progenitor itself ({\it self}-enrichment models).  In the
second class of models, the issue of their formation is directly related 
to the origin of their metal content (Cayrel 1986; Brown, 
Burkert \& Truran 1995; Parmentier et al.~1999).
Such a feature makes the self-enrichment scenario especially appealing if 
we hypothesize that the GC progenitor clouds are made of 
primordial (i.e. metal-free) gas, a reasonable assumption for the 
Old Halo GCs, i.e. the population of old and coeval (Rosenberg et al.~1999) 
halo GCs.

Regarding the origin of the proto-globular cluster clouds (PGCC), 
Fall \& Rees (1985) suggested that they formed out of the collapsing 
protoGalaxy, as cold and dense clouds in pressure equilibrium with a 
hot and diffuse background.  In the frame of this theory, the GC 
progenitor clouds are thermally supported (i.e. no additional support 
against gravitation due to magnetic fields and turbulence) and made of 
primordial gas.  In order to explain the metallicities of halo GCs, 
Parmentier et al.~(1999, hereafter Paper I) expanded the Fall \& Rees (1985)
model for GC formation by the self-enrichment picture. 
According to this one, a first stellar generation forms in 
the central regions of each PGCC.  When the massive stars explode
as Type II supernovae (SNeII), they chemically enrich the surrounding gas
and sweep the cloud, turning it in an expanding shell of gas in which
the formation of a second, chemically enriched, stellar generation may
be triggered.  These second generation stars form the 
proto-globular cluster. \\

The sweeping and compression of the interstellar medium by massive star 
explosions is not the sole mechanism invoked to account for star formation 
during the earliest stages of the Galactic evolution.  
Some other models assume a different origin for the trigger.
For instance, following Vietri \& Pesce (1995), the high pressure 
confining the GC gaseous progenitor leads to the propagation of
a strong shock inwards the cloud, stimulating thereby the formation of 
new stars.  On another side, Murray \& Lin (1992) and Dinge (1997) 
suggested that the propagation of shock waves is promoted by cloud-cloud 
collisions.  In a similar way, star forming clouds may have coalesced
into larger units until they reach a density and/or accumulated 
mass high enough to enable the formation of bound globular clusters 
(Larson 1988, Smith 1999).  Obviously, several
processes are able to collect ambient gas into dense layers/clouds 
where star formation can thereafter take place.  Several of them may  
have been at work at the same time.  Our interest being in understanding
the origin of the metal content of GCs, in what follows, we investigate
the hypothesis of star formation in gas layers swept by the  
explosions of PopIII massive stars located at the center of the PGCCs. \\

The debate of how PopIII stars looked like has been raging for many years.
Numerous studies have addressed the issue of the collapse and fragmentation of
primordial gas clouds in order to estimate the masses of PopIII stars.  
The primordial gas being deficient in heavy elements (i.e., the most 
efficient coolants in present-day star forming clouds), all of them
have emphasized the importance of cooling by H$_2$ molecules.
Inspite of this, the achieved conclusions do not necessarily converge.
Recent numerical simulations (Abel, Bryan \& Norman 2002) suggest that
metal-free stars form in isolation and are massive 
(30 $\lesssim$ M $\lesssim$ 100~M$_{\odot}$) objects.  From their own 
simulations, Bromm, Coppi \& Larsson (1999)
quoted a characteristic mass even larger than 100~M$_{\odot}$.  These
results are in marked contrast with the early study performed by Palla, 
Salpeter \& Stahler (1983) following which primordial gas clouds should
be capable of fragmenting into low-mass stars 
(i.e., down to $\sim$ 0.1~M$_{\odot}$).
Nakamura \& Umemura (1999) reached a somewhat intermediate conclusion.
According to them, the mass range of the first stars is similar 
to its present value (i.e. no star with M $\gtrsim$ 100~M$_{\odot}$) but 
 nevertheless excludes low-mass long-lived stars, that is, the lowest
mass star allowed to form in a metal-free medium is a $\simeq$ 3~M$_{\odot}$ 
star.  In a more recent model, Nakamura \& Umemura (2001) 
predict a bimodal initial mass function for PopIII stars.  In fact,
the initial mass function of metal-free stars would show two distinct
peaks at $\simeq$ 1~M$_{\odot}$ and $\simeq$ 100~M$_{\odot}$, that is, it 
would include intermediate mass (1 $\lesssim$ M $\lesssim$ 10~M$_{\odot}$)
stars, massive (10 $\lesssim$ M $\lesssim$ 100~M$_{\odot}$) stars
as well as very massive (M $\gtrsim$ 100~M$_{\odot}$) ones.
As quoted by Christlieb et al.~(2003), the recent discovery of HE 0107-5240,
the most metal-poor ([Fe/H]=$-$5.3) star ever discovered in the Galactic halo,
could be a challenge to these models precluding the formation of stars
with mass low enough for their life duration to exceed a Hubble time.

While there is currently a wide consensus that present star formation
operates predominantly in a clustered mode (Lada, 
Strom \& Myers 1993), the question whether clusters of
metal-free stars managed to form in the first (proto-) galaxies is still
open.  In this paper, we assume that at least some of the primordial 
star formation sites produced stellar {\it clusters}
whose most massive stars (10 $\lesssim$ M $\lesssim$ 50\,M$_{\odot}$) 
end their life 
as canonical SNeII.  Therefore, our study does not include 
the possibility of a prompt enrichment of the primordial interstellar
medium by a population of very massive (i.e., $\gtrsim$ 100~M$_{\odot}$) 
stars as suggested by, e.g., 
Wasserburg \& Qian (2000).  The chemical enrichment provided by isolated
(very) massive objects will be qualitatively discussed in Sect.~3
where we will see that they may be appropriate to account for the
formation of very metal-poor stars, i.e., stars more metal-poor than the 
most metal-deficient halo GCs ([Fe/H] $\lesssim$ $-$2.5).  \\
      
Supernovae having long been thought to disrupt the cloud of gas out of 
which they have formed, Parmentier et al.~(1999) studied the ability of 
pressure-truncated clouds to retain SNII ejecta.  This ability will rule 
the metal content of the proto-cluster.  In fact, in this class of models,
the final metallicity is determined by the number of supernovae 
($N$, which, assuming a given initial mass function and given SNII yields, 
determines the amount of metals dispersed within the PGCC) and 
the background pressure ($P_h$, which determines the mass of the cloud, 
that is, the mass of primordial gas to be chemically enriched).
Comparing the gravitational energy of the PGCC with the kinetic energy of 
the supershell resulting from the SNII explosions, Parmentier et al.(1999) 
showed that the gaseous GC progenitors can sustain up to $\sim$ 200 supernovae 
(the disruption criterion, Eq.~14, Paper I).  This is a number high enough 
for the PGCCs to achieve halo metallicities.  Furthermore, for a given 
number of exploding massive stars, a self-enrichment episode in 
pressure-bound clouds lead to a metallicity gradient throughout the 
resulting system of GCs and to a correlation between the mass and the 
achieved metallicity of the gaseous progenitors in the sense that the 
least massive clouds are the most metal-rich.  Such a trend emerges 
because if the bound pressure is higher, the mass of the pressure-truncated 
cloud will be lower, and its ability to retain supernova ejecta will be 
greater.  These trends, i.e. a metallicity gradient and a 
mass-metallicity relation, are indeed statistically present in the Old Halo, 
that is, the halo from which the presumably accreted component has been 
removed (Parmentier et al.~2000, Parmentier \& Gilmore 2001). \\

After having analysed to which extent the Galactic halo GC data fit the 
correlations induced by the self-enrichment process, the next step is 
to wonder whether there are some stars forming out of the chemically 
enriched supershell, i.e. whether there is a second stellar generation 
tracing these correlations. 
The idea of star formation induced by supernova explosions dates back at 
least to Opik (1953).  Numerous cases of distant star forming loops,
connected with shells resulting from supernova explosions and located
in the Galactic disc as well as in dwarf galaxies, are detailed
in the literature (e.g. Comeron \& Torra 1994, Walter et al. 1998, 
Efremov \& Elmegreen 1998, Rubio et al. 1998).  In this paper, we address
the specific case of supershells made of swept PGCCs and 
within which the formation of the stars of future GCs may be triggered.
In a first step, we limit our study to the propagation of the shell
throughout the hot protogalactic background in which the PGCCs are 
initially embedded, this part of the shell propagation being much longer than 
the propagation throughout the cloud (see Fig.~\ref{fig:RsHPB}).  \\ 

The outline of the paper is as follows.  In Sect.~2, we solve the perturbed 
equations of continuity and motion for transverse flows within the shell
(i.e. the swept cloud) in order to identify the conditions supporting a 
successful shell transverse collapse.  We also discuss in turn the impact of 
the different parameters acting upon the shell collapse and, thus, upon the 
temporal growth of the shell fragments in which further star formation 
may be stimulated.  In Sect.~3, we show how the conditions required to 
stimulate a star formation episode within the shell provides a natural 
explanation to the observed 
metallicity range of halo GCs and how the shape of their metallicity spectrum
constitutes the next step to work on.  Sect.~4 describes some effects
which our forthcoming computations should take into account in order to 
refine the present model.  Finally, our conclusions are presented in Sect.~5.

\section[]{Stimulated Star Formation in Proto-Globular Cluster Clouds}
\label{sec:stimSF}
The shell will in general contain perturbed (transverse) velocity 
components and perturbations in the column density whose development
leads to the transverse collapse of the swept PGCC and, thereby, to
the formation of a second stellar generation.  We now derive the conditions
to get such a collapsing shell.  The elementary method described below 
is adopted.
\subsection{Modelling the transverse collapse of the shell}
The computations are based on the linear perturbed equations of continuity 
and motion for transverse flows in the shell (e.g. Elmegreen 1994).  \\
The perturbed equation of continuity (mass conservation) is

\begin{equation}
\frac{\partial \sigma _1}{\partial t} = -2 \frac{V_s}{R_s} \sigma _1
- \sigma _0 \nabla _T . v\,,
\label{eq:per_cont}
\end{equation} 

where subscript T means that the gradient
component under consideration is the transverse one,
and the perturbed equation of motion (momentum conservation) is
\begin{equation}
\sigma _0 \frac{\partial v}{\partial t} = - \sigma _0 \frac{V_s}{R_s} v
- {c_s}^2~ \nabla \sigma _1 + \sigma _0 g_1\;.
\label{eq:per_motion}
\end{equation}  

In these equations, $R_s$ and $V_s$ are respectively the radius and the
velocity of the shell, $\sigma _0$ is the unperturbed surface
density, $\sigma _1$ is the perturbed surface density, $v$ is the
perturbed (transverse) velocity, $c_s$ is the velocity dispersion 
of the material inside the shell, $g_1$ is the perturbed gravity, this one
being related to the surface density through (Elmegreen 1994):
\begin{equation}
g _1 = -2 \pi i G \sigma _1\;.
\label{g1_sigma1}
\end{equation} 
As mentioned above, the evolution of the shell is studied while propagating
through the hot background, i.e. when the whole cloud has been swept 
inside the shell. 
The hot background being a diffuse medium, the shell mass $M_s(t)$ does not 
increase any longer at this stage of its propagation and is given by the mass 
$M$ of the progenitor cloud.
The unperturbed surface density of the shell is thus given by
\begin{equation}
\sigma _0 = \frac{1}{4 \pi} \frac{M_s(t)}{R_s ^2} 
= \frac{1}{4 \pi} \frac{M}{R_s ^2}\;. 
\label{sigma_0}
\end{equation}

Equation \ref{eq:per_cont} shows that the development with time
of any perturbation of the shell surface density  
($\partial \sigma _1 / \partial t  > 0$) is inhibited by the 
stretching of the perturbed region
due to the shell expansion (i.e. $V_s > 0$, first term on the right 
hand-side, hereafter rhs) while the convergence of the perturbed flows 
supports the growth of the perturbation (second term on the rhs).    
Equation \ref{eq:per_motion} shows that an initial transverse flow 
of material along the shell develops ($\partial v/ \partial t  > 0$)
only if the self-gravity (third term on the rhs) overcomes
the stabilizing effects of the stretching (first term on the rhs)
and of the internal pressure (second term on the rhs), here represented
by $c_s^2$, the shell sound speed squared. \\

In order to solve Eqs.\ref{eq:per_cont} and \ref{eq:per_motion} properly,
we now turn to the determination of the expansion law of the shell, 
i.e. $R_s(t)$ and $V_s(t)$, while it propagates throughout the hot 
protogalactic background.

\subsection{Supershell propagation throughout the hot background}
\label{sub:propa_HPB}

\begin{figure}
\begin{center}
\epsfig{figure=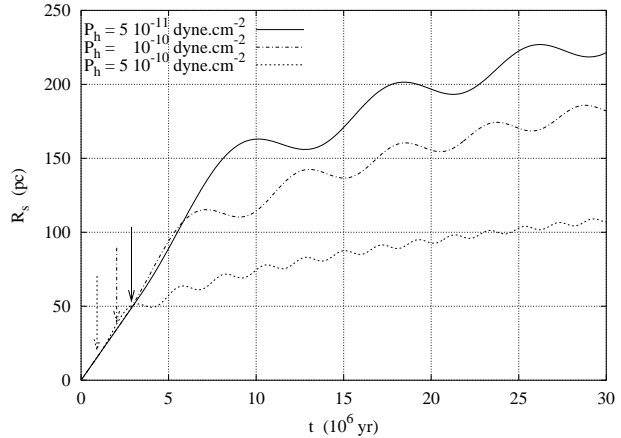, width=\linewidth}
\caption{Evolution with time of the shell radius inferred from 
Eqs.~(\ref{eq:bubble_nrj_HPB} - \ref{eq:mass_sh_HPB}) assuming that 200\,SNeII
explode at a constant rate during 30 million years and for three 
different hot protogalactic background pressures, from top to bottom 
5$\times 10^{-11}$ (plain curve), 
10$^{-10}$ (dashed-dotted curve) and 5$\times 10^{-10}$\,dyne.cm$^{-2}$ 
(dotted curve).  In each case, an arrow indicates the time $t_{em}$ at which 
the shell crosses the interface between the cold and hot phases} 
\label{fig:RsHPB}
\end{center}
\end{figure}

The equations describing the propagation of a supernova-driven shell
are as follow (Castor et al.~1975). 
\begin{enumerate}
\item The supernova explosions add energy to the bubble at a constant rate
$\dot{E_o}$ and the dominant energy loss of the bubble comes from the work
against the dense shell, hence the variation with time of the energy 
$E_b$ of the bubble obeys
\begin{equation}
\dot{E_b}=\dot{E_o}-4\pi {R_s}^2 P_b \dot{R_s}\,.
\label{eq:bubble_nrj_HPB}
\end{equation}

We assume that the kinetic energy of every SN is 10$^{51}$~ergs and that 
the SN phase lasts about thirty million years, 
i.e. $\dot{E_o}=N 10^{51} {\rm ergs}/30 {\rm Myr}$.

\item The internal energy $E_b$ and the pressure $P_b$ of the bubble are 
related through 
\begin{equation}
\frac{4\pi}{3} {R_s}^3 P_b = \frac{2}{3} E_b\,,
\label{eq:bubble_P_HPB}  
\end{equation}

\item The shell motion obeys Newton's second law 
\begin{equation}
\frac{d}{dt} [M_s(t) \dot{R_s}(t)] = 4\pi {R_s}^2 (P_b-P_{ext})
-\frac{{GM_s}^2(t)}{2{R_s}^2(t)}\,,
\label{eq:Newton2_HPB}  
\end{equation}
where $P_{ext}$ is the pressure of the medium just outside the shell.
\item Considering the case of swept PGCCs propagating through the hot 
protogalactic background, the mass of the shell is constant in time 
and is given by the mass of the cloud (see Sect.~2.1):
\begin{equation}
M_s(t) = M\;. \\
\label{eq:mass_sh_HPB}
\end{equation} 
\end{enumerate}

The pressure external to the shell is exerted by the surrounding hot 
background (i.e. P$_{ext}$ = P$_h$ in Eq.~\ref{eq:Newton2_HPB}) and is 
therefore assumed to be constant in time.  This pressure
will strongly decelerate the shell as illustrated below.
Numerical resolution of Eqs.~(\ref{eq:bubble_nrj_HPB}
- \ref{eq:mass_sh_HPB}) provides $R_s(t)$ and thereby $V_s(t)$ 
and $\sigma _0(t)$.  The initial conditions are those at the time $t_{em}$,
i.e. when the shell crosses the interface between the cloud and the 
surrounding background: the mass and radius of the shell are those of the 
pressure-bound cloud ($R_s(t_{em})=R$, $M_s(t_{em})=M$), the velocity of 
the shell is determined by its former propagation at constant speed $V$ 
through the cloud (see Paper I, Eq.~13), that is $V_s=V$ and $A_s=0$. \\

\begin{figure}
\begin{center}
\epsfig{figure=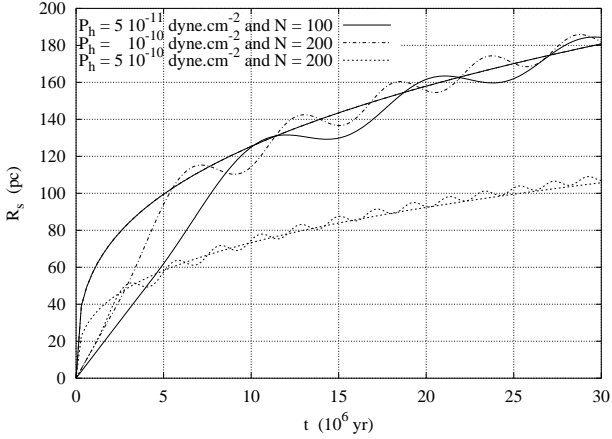, width=\linewidth}
\caption{Comparison between the evolution with time of the shell radius 
computed from Eqs.~(\ref{eq:bubble_nrj_HPB} - \ref{eq:mass_sh_HPB})  and
the average radius given by Eq.~9 (i.e. $<R_s(t)> \propto t^{1/3}$) 
for the quoted numbers $N$ of SNeII and background pressures $P_h$}
\label{fig:RsHPBt1/3}
\end{center}
\end{figure}

Figure \ref{fig:RsHPB} shows the evolution with time of the shell radius 
for 200\,SNeII, i.e. the maximum number of supernovae that a PGCC can sustain 
(disruption criterion, Paper I), and for 3 different values of the hot 
protogalactic background pressure, i.e. $P_h$ = 5$\times 10^{-11}$, 
10$^{-10}$ and 5$\times 10^{-10}$\,dyne.cm$^{-2}$.
The SN rate being the same in the three cases, the shells propagate 
at the same velocity in the cold phase (Eq.~13 in Paper I).  Their 
expansions begin to differ once they have crossed the interface between 
their respective PGCC and the hot background in which the latter is embedded, 
this time being indicated by an arrow in Fig.~\ref{fig:RsHPB}.  Obviously, 
the propagation of the shell through the cloud is much shorter than its 
propagation through the background.   A significant part of the shell 
expansion through the background takes place at early time.  Indeed,
after a transient phase during which the velocity of the shell does not differ 
markedly from its velocity inside the cloud (Fig.~\ref{fig:RsHPB}), 
the overall expansion slows down and the radius of the shell scales 
roughly as $t^{1/3}$. 
This average expansion can be obtained from Eqs.~(\ref{eq:bubble_nrj_HPB}
- \ref{eq:mass_sh_HPB}) assuming that $M_s(t)=0$, reflecting thereby that the 
temporal evolution of the shell radius does not depend strongly on the mass 
(Brown et al.~1995):
 
\begin{equation}
<R_s(t)> = \left (\frac{3}{10\pi} \frac{\dot{E_o}}{P_h} \right )^{1/3} t^{1/3}.
\label{eq:RsHPBt1/3}
\end{equation} 

Figure \ref{fig:RsHPBt1/3} shows the good agreement between 
Eq.~\ref{eq:RsHPBt1/3} and the result of the numerical integration over 
time of Eqs.~(\ref{eq:bubble_nrj_HPB} - \ref{eq:mass_sh_HPB}).  
Equation \ref{eq:RsHPBt1/3} shows that, during the long-term evolution,
the expansion rates of two shells having the same $(N/P_h)$ ratio are similar.
Figure \ref{fig:RsHPBt1/3} illustrates this effect for 
two sets of values, namely $N$=100 and $P_h$=5$\times 10^{-11}$\,dyne.cm$^{-2}$
(plain curve) and $N$=200 and $P_h$=$10^{-10}$\,dyne.cm$^{-2}$  
(dashed-dotted curve). 

\subsection{Growth with time of the shell fragments}
\label{sec:frag_num}

In order to assess whether the shell transverse collapse proceeds successfully
or not, we now numerically integrate over time Eqs.\ref{eq:per_cont} and 
\ref{eq:per_motion} in order to derive the temporal evolutions of the 
perturbed and unperturbed surface densities,
$\tilde{\sigma} _1(t)$ and $\sigma _0(t)$, respectively. 

Assuming that the perturbed quantities follow a complex exponential of the 
angular position $\phi$ along the shell, we get:
\begin{equation}
\sigma _1 (t,\phi) = \tilde{\sigma} _1 (t) ~e^{-i \eta \phi}
\label{sigma1_exp}
\end{equation}
and
\begin{equation}
v (t,\phi) = \tilde{v} (t)  ~e^{-i \eta \phi} ~e^{i \Delta \phi}\,.
\label{v_exp}
\end{equation}

In these equations, $\Delta \phi$ represents the phase difference between 
the perturbed surface density $\sigma _1$ and the perturbed velocity $v$.  
$\eta$ is the angular wavenumber and is related to the spatial wavenumber $k$ by:
\begin{equation}
\eta = k R_s = \frac{2 \pi}{\lambda}~R_s 
\label{eta_lambda}
\end{equation}
where $\lambda$ is the wavelength of the perturbation, namely the 
average distance between forming fragments, the sites of future star 
formation.  Therefore, $\eta$ is the number of forming 
clumps along a shell circumference and, as such, $\eta$ must be an integer.
Moreover, any realistic perturbation must fit inside a fraction
of the shell circumference, say, $\lambda \leq R_s$ or $\eta \geq 6$.  \\

At a time $t$ and an angular position $\phi$ along the shell,
the shell surface density $\sigma _s$ obeys
\begin{equation}
\sigma _s(t,\phi )=\sigma _0(t)+\sigma _1(t,\phi )
                  =\sigma _0(t)+\tilde{\sigma _1}(t)~cos(\eta \phi)\;.
\label{eq:sigt_t_anlt}
\end{equation} 

The fragmentation of the shell, i.e.~its fully developed transverse collapse, 
occurs when
\begin{equation}
\tilde{\sigma} _1(t) = \sigma _0(t). 
\end{equation} 

Writing $- i \eta /R_s$ for the gradient and using 
Eqs.~\ref{g1_sigma1} and \ref{sigma_0}, 
Eqs.~\ref{eq:per_cont} and \ref{eq:per_motion} successively become:

\begin{itemize}
\item[$\triangleright$] Perturbed equation of continuity: 

\begin{eqnarray}
\frac{\partial \tilde{\sigma _1}}{\partial t} 
& = & -2\frac{V_s}{R_s} ~\tilde {\sigma _1} 
 +  \sigma _0 ~ \frac{i~\eta}{R_s} ~ \tilde{v}~e^{i\Delta \phi} \nonumber \\
\frac{\partial \tilde{\sigma _1}}{\partial t}
& = & -2\frac{V_s}{R_s} ~\tilde{\sigma _1} 
 +  \frac{i~\eta~M}{4 \pi} \frac{\tilde{v}}{{R_s}^3} ~e^{i\Delta \phi} 
\label{eq:per_cont_num}
\end{eqnarray}

\item[$\triangleright$] Perturbed equation of motion:

\begin{eqnarray}
\sigma _0 ~e^{i\Delta \phi} ~\frac{\partial \tilde{v}}{\partial t}  
& = & -\sigma _0 ~ \frac{V_s}{R_s} ~ \tilde{v}~e^{i\Delta \phi} 
 + {c_s}^2 ~ \frac{i~\eta}{R_s} ~ \tilde{\sigma _1} \nonumber \\
 & & \,-  2 \pi i G ~ \sigma _0 ~ \tilde{\sigma _1} ~~~ \nonumber \\
~\frac{\partial \tilde{v}}{\partial t}  
& = & - \frac{V_s}{R_s} ~ \tilde{v} 
\hspace{1.3cm} \,+ \frac{4~\pi ~i ~\eta ~c_s^2}{M} R_s ~\tilde{\sigma _1} 
~ e^{- i\Delta \phi} \nonumber \\
 & & -  2 \pi i G \tilde{\sigma _1} e^{- i\Delta \phi} ~~~
\label{eq:per_motion_num}
\vspace*{3pt}
\end{eqnarray} 
\end{itemize}

Seven parameters intervene in the ability of the shell
to undergo a transverse collapse:
\begin{itemize}
\item[- ] the supernova number $N$ and the pressure external 
to the shell $P_h$, which determine the shell expansion law ($R_s$ depends
on $N$ and $P_h$) and mass ($M$ depends on $P_h$),
\item[- ] the initial values of the perturbed quantities, 
$\tilde \sigma _1(t_{em})$ and $\tilde v (t_{em})$, 
and the associated phase difference $\Delta \phi$,
\item[- ] the number $\eta$ of forming clumps embedded along a shell
circumference,
\item[- ] the velocity dispersion $c_s$ of the shell material which is 
related to the internal pressure $P_s$ of the shell, supporting it against 
transverse collapse, through $P_s={c_s}^2 \rho _s$ ($\rho _s$ is the
volumic mass density of the shell material).  
\end{itemize}

In what follows, the influence of each of the parameters involved in the 
fragmentation process is investigated, the final aim being to check whether 
some reasonable sets of conditions can lead to the shell fragmentation.   

\subsubsection{The phase difference between $v$ and $\sigma_1$: $\Delta \phi$}
\label{sub:delta_phi}

\begin{figure}
\hspace*{5mm}
\epsfig{figure=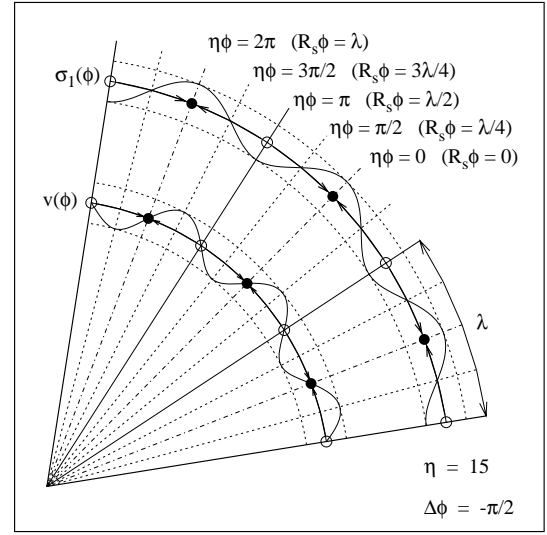, width=11cm}
\caption{Schematic description of the sinusoidal (i.e. complex exponential) 
behaviour with $\phi$ of the perturbed quantities $v$ (lower curve)
and $\sigma _1$ (upper curve) along a fraction of the shell circumference 
at a given time assuming that the number of clumps $\eta$ is 15 and 
the phase difference $\Delta \phi$ between $v$ and $\sigma _1$ is 
$-\frac{\pi}{2}$.  Such a phase difference corresponds to the convergence 
of the transverse (perturbed) flows (indicated by the thick arrows) 
towards the initial clumps (filled circles).  
The open circles represent the shell regions which 
are progressively depleted by the transverse flows}
\label{fig:fragshell}
\end{figure}

The amplitude of the perturbed surface density at a time $t$,  
$\tilde \sigma_1(t)$, results from its initial value
$\tilde \sigma_1(t_{em})$ and from the transverse flows which 
redistribute the shell mass accumulated while the shell was propagating 
throughout the progenitor cloud.  The fragmentation 
is therefore favoured if these transverse flows 
converge towards the clumps initially present within the shell.  
Such a situation corresponds to a phase difference 
$\Delta \phi$ = $-\frac{\pi}{2}$ between the two perturbed quantities
$\sigma _1(t,\phi)$ and $v(t,\phi)$, that is,  the transverse velocity 
exhibits a phase delay of a quarter of a wavelength with respect to the 
perturbed surface density (see Fig.~\ref{fig:fragshell}). 
Replacing $\Delta \phi$ by this value, 
Eqs.~\ref{eq:per_cont_num} and \ref{eq:per_motion_num} respectively become:

\begin{equation}
\frac{\partial \tilde{\sigma _1}}{\partial t}=-2\frac{V_s}{R_s} ~\tilde
{\sigma _1} + \frac{\eta~M}{4 \pi} \frac{\tilde{v}}{{R_s}^3}  
\label{eq:per_cont_num_phi}\,,
\end{equation}

\begin{equation}
\frac{\partial \tilde{v}}{\partial t}
=-~\frac{V_s}{R_s} ~ \tilde{v} 
- \frac{4~\pi  ~\eta ~c_s^2}{M} R_s ~\tilde{\sigma _1} 
+ 2\pi G ~ \tilde{\sigma _1}\;.  \\
\label{eq:per_motion_num_phi} 
\end{equation}

In what follows, the roles of the other parameters is studied 
assuming that $\Delta \phi = - \pi $/2, i.e. through the numerical integration
of Eqs.~\ref{eq:per_cont_num_phi} - \ref{eq:per_motion_num_phi}. 

\subsubsection{The global parameters: $N$ and $P_h$}
\label{sub:discNPh}

\begin{figure}
\epsfig{figure=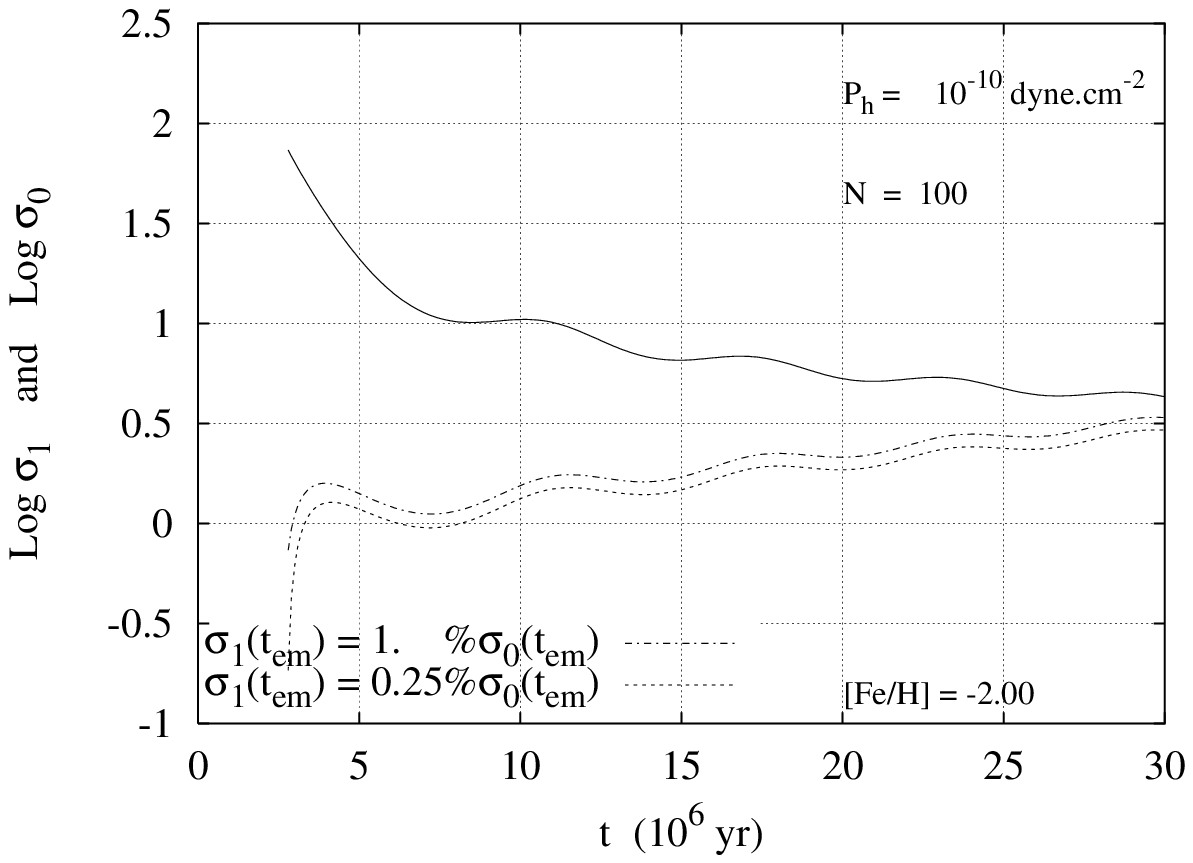, width=\linewidth}
\epsfig{figure=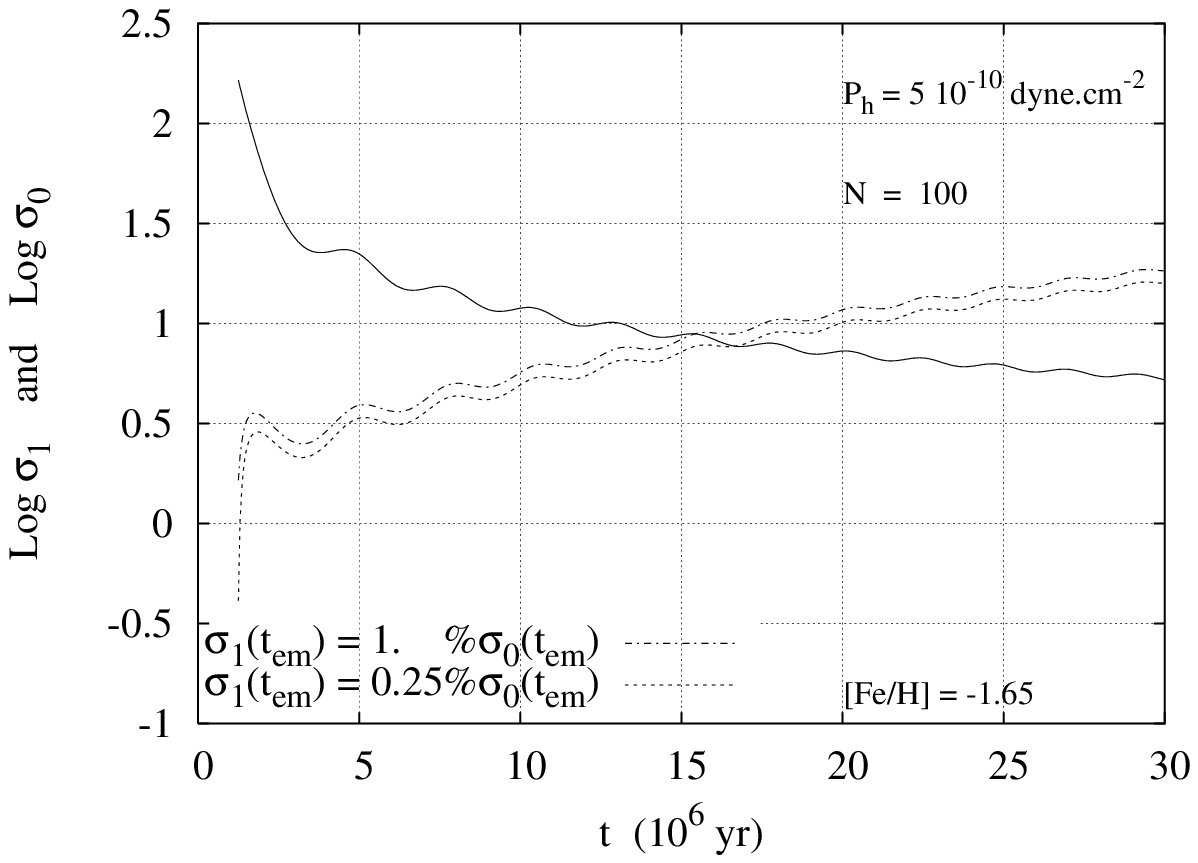, width=\linewidth}
\epsfig{figure=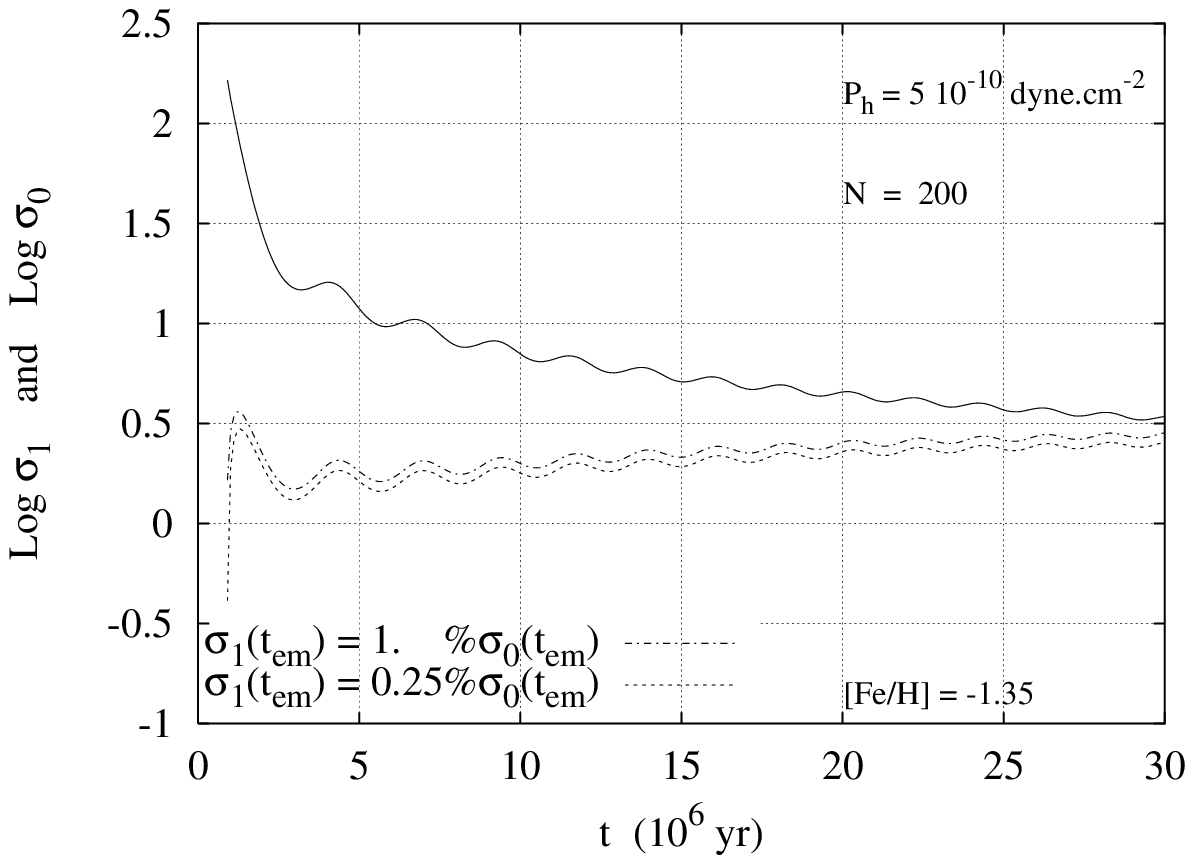, width=\linewidth}
\caption{Temporal evolution of $\sigma _0$ (plain curves) and 
$\tilde \sigma _1$ (dashed-dotted curves), expressed in units of 
1\,M$_{\odot}$.pc$^{-2}$, considering 
two different initial perturbed surface densities (see the keys)
and the external pressures and SNII numbers indicated.  In each panel, 
$c_s$=1\,km.s$^{-1}$, $\eta=10$ and $v(t_{em})=0.01\,V_s(t_{em})$}
\vspace{11pt}
\vspace{11pt}
\label{fig:disc_sig1init}
\end{figure}

In order to assess whether the shell fragments or not during its propagation
through the hot protogalactic background, we compare in 
Fig.~\ref{fig:disc_sig1init} the evolutions with time of the perturbed 
and unperturbed surface densities, namely $\tilde{\sigma} _1(t)$ and 
$\sigma _0(t)$, for different values of the external pressure $P_h$ 
(5$\times 10^{-10}$ and 10$^{-10}$\,dyne.cm$^{-2}$)
and numbers $N$ of SNeII (100 and 200).  
The corresponding metallicity is indicated in each panel.  
The other parameters are kept the same in every case: 
the sound speed $c_s$ of the shell material is 1\,km.s$^{-1}$, 
the perturbed velocity $v(t_{em})$ is 0.01\,$V_s(t_{em})$, the
number of forming clumps $\eta$ is 10 and the initial perturbed 
surface density is assumed to be of order one per cent
of the unperturbed value, namely 
$\sigma _1(t_{em}) = 0.01 \times \sigma _0(t_{em})$.
This choice may seem rather arbitrary but the next paragraph will show that 
the initial amplitude of  $\sigma _1$ does not affect the 
results significantly .  \\

While the unperturbed surface density decreases due to the shell 
expansion at constant mass, the perturbed surface density grows
at a rate which depends on the external pressure and on the number of SNeII. 
The comparison of the different panels in Fig.~\ref{fig:disc_sig1init} 
shows that the development of $\tilde{\sigma} _1(t)$ with respect to 
$\sigma_0(t)$ is favoured by a {\sl a high background pressure} 
and a {\sl low number of SNeII}.  
Among the three panels presented ($P_h=10^{-10}$\,dyne.cm$^{-2}$ and $N$=100,
$P_h=5\times 10^{-10}$\,dyne.cm$^{-2}$ and $N$=100,
$P_h=5\times 10^{-10}$\,dyne.cm$^{-2}$ and $N$=200), the transverse 
collapse of the shell is successfully achieved\footnote{At this point, 
it should be kept in mind that once the shell has achieved its complete 
transverse collapse, Eqs.~\ref{eq:per_cont} and Eqs.~\ref{eq:per_motion}
are no longer valid.  Therefore, once ${\sigma} _0 \leq \tilde\sigma _1$, 
the curve $\tilde \sigma _1(t)$ has no longer physical meaning.} 
when the lower number of supernovae is combined with the larger external 
pressure 
($P_h=5\times 10^{-10}$\,dyne.cm$^{-2}$ and $N$=100).
This effect comes from the dependence of the shell surface density on 
the external pressure $P_h$ and the supernova number $N$.  Indeed,  
combining the mass of a pressure-truncated gas cloud 
(i.e. $M \propto P_h^{-1/2}$) with Eq.~\ref{eq:RsHPBt1/3}, the 
surface density of the shell scales as:
\begin{equation}
\sigma _0  (t) \propto \frac{M}{R_s(t)^2} \propto \frac{P_h^{1/6}}{N^{2/3}}.
\label{eq:sig0_N_Ph}
\end{equation} 
Consequently, the larger the pressure and/or the lower the supernova number,
the larger the shell surface density and the larger the growth rate of 
the perturbation (see Eq.~\ref{eq:per_cont} and Eq.~\ref{eq:per_motion}, 
a larger surface density favours the terms promoting an efficient
transverse collapse).
{\sl Therefore, although the presence of exploding massive stars is required
to stimulate the formation of a second stellar generation,
too large a number of SNeII inhibits the ability of the shell to 
collapse and to form new stars}\footnote{Unlike the protogalactic shells 
we study, the collapse of shells expanding in the Galactic disc is made 
easier by a larger number of SNeII.  In fact, assuming that the surrounding
interstellar medium is roughly homogeneous, i.e. the shell radius is smaller 
than the density scale-height of the Galactic HI layer, one gets by mass 
conservation, assuming a pre-shell density 
$\rho _0$, $\sigma _0 = \frac{R_s \rho _0}{3}$.  In this case, a larger 
number of SNeII leads to a larger radius at a given time and, therefore, 
to larger surface density and perturbation growth rate.}.  
 
The influence of the surface density of the shell on its collapse 
is reminiscent of the star formation law 
on large scales, i.e. averaged over entire galactic discs.  In that case,
providing that the gas surface density is larger than a density 
threshold, the star formation rate follows a power-law of the gas 
surface density (the so-called Schmidt law) while it falls sharply 
below (Kennicutt 1989).  The threshold surface
density varies from one galaxy to another but remains nevertheless in the 
range 10$^{20}$-- 10$^{21}$\,cm$^{-2}$.  It may not be a coincidence that
the shell surface densities obtained in the frame of this model are of 
the same order of magnitude or even larger (see Sect.~2.3.3).

The dependence of the perturbation growth rate on $N$ and $P_h$ 
raises a point of interest, worthy of a mention here.
It just so happens that the parameters determining the final metallicity 
of the proto-cluster, namely the number of supernovae ($N$, which determines
the amount of metals dispersed within the PGCC) and the background pressure 
($P_h$, which determines the mass of the cloud, that is, the mass of primordial
gas to be chemically enriched)
are also some of those influencing the ability of the shell to form new
stars.  {\sl If some combinations of external pressures and SN numbers support 
the completion of the shell transverse collapse much more than some others 
do, then the formation of second stellar generations with the corresponding 
metallicities will be favoured}.  Therefore, in the frame of the 
self-enrichment scenario, there is a direct link between the achieved 
metallicity and the probability of forming halo stars, i.e. {\sl the 
study of the fragmentation process may shed light on the metallicity 
distribution function of Galactic halo field stars and GCs}.    

Figure \ref{fig:disc_sig1init} and Eq.~19 show that the shell fragmentation 
is favoured by a low number of SNe and a high pressure of the hot 
protogalactic background.  Now, let us imagine that, on the contrary,
the collapse efficiency increases with both decreasing number of SNeII 
and pressure.  As $P_h$ and $N$ get smaller, the number of successful 
transverse collapses increases and the newly formed stars are more metal-poor.
Thus, the metallicity distribution function
of the Galactic halo would exhibit an increasing number of
proto-clusters/halo stars with decreasing metallicity.  If, on the other 
hand, the shell ability to achieve fragmentation increased with
both increasing external pressure and SN number, then the larger $P_h$ 
and $N$, the more numerous the transverse collapses and the more 
metal-rich the newly formed stars.  As a result, there would be 
an increasing number of proto-clusters/halo stars with increasing metallicity.
Neither an increasing nor a decreasing metallicity distribution
function is observed for the Galactic halo.  In contrast, the
metallicity distributions, for both halo field stars 
(Laird et al.~1988) and halo GCs (Zinn 1985),
are peaked-shape.  While the finding that the shell transverse collapse 
is favoured by large external pressures (promoting ``large'', i.e. mildly 
metal-poor, metallicities)
and low SN numbers (supporting low metallicities) is not sufficient 
in itself to draw some definitive conclusions regarding the shape
of the halo metallicity distribution, it appears that it does not 
contradict it either.  

\subsubsection{The initial perturbed surface density: $\tilde{\sigma} _1(t_{em})$}
\label{sub:disc_sig1init}

Protogalactic shells such as those studied here exhibit number surface
densities of order several 10$^{20}$\,cm$^{-2}$.
Indeed, using Eqs.~\ref{sigma_0} and \ref{eq:RsHPBt1/3}, one gets:
\begin{eqnarray}
\sigma _0(t, N, P_h) & \simeq & 5 \times 10^{-3} P_{h(10)}^{1/6} 
N_{200}^{-2/3} t_6^{-2/3}\, {\rm g.cm^{-2}} \nonumber \\
 & \simeq & 2.5 \times 10^{21} P_{h(10)}^{1/6} 
N_{200}^{-2/3} t_6^{-2/3}\,{\rm cm^{-2}}
\end{eqnarray}

where the subscript (10) means that the pressure is expressed in units of
$10^{-10}\,\rm dyne.cm^{-2}$, the subscript 200 means that $N$ is expressed 
in units of 200\,SNe and the subscript 6 means that the time is 
expressed in units of $10^6$\,years.  Considering the interstellar medium
of the Galactic disc with roughly the same surface density (though a bit 
lower, i.e. $\simeq$ 10$^{20}$\,cm$^{-2}$), the relative non-uniformities 
in the number surface density are typically of order 0.01 - 0.2 
(Wunsch \& Palous 2001).  However, the initial perturbation in the 
shell surface density is certainly not as large as some of the clumps of this 
{\sl quiescent} disc interstellar medium since {\sl turbulent mixing} is expected to 
take place within the shell (this is an important requirement to achieve 
the efficient mixing of the supernova heavy elements with the cloud gas
and, therefore, to maintain the chemical homogeneity of the shell, that is,
of the proto-globular cluster; see Brown et al.~1991), thus lowering 
the surface density inhomogeneities. \\

Figure 4 also shows the temporal evolution of $\sigma _0$ and 
$\tilde \sigma _1$ assuming that the initial perturbed surface density is 
lower than in Sect.~2.3.2, namely 0.25\%\,$\sigma _0(t_{em})$ instead of 
1\%\,$\sigma _0(t_{em})$.
Despite these different initial values, each panel displays almost 
identical $\tilde \sigma _1(t)$ curves in both cases.  Owing to the initial 
transverse flows, the initial clumps quickly undergo a 
replenishment in shell material leading afterwards to very similar temporal
evolutions of the perturbed surface density whatever the initial perturbed 
surface density.  Obviously, this one is not a key parameter 
of the fragmentation process. 

\subsubsection{The initial perturbed velocity: $v(t_{em})$}
\label{sub:disc_vtrans}

All the results presented here above assume the spherical symmetry of the
supershell.  It is clear however that such a system is
not expected to remain perfectly spherical.  The deviations from spherical
symmetry can arise, for instance, from the non point-like nature
of the energy input, namely all the stars of the first generation
cluster are not located exactly at the centre of the cloud/shell.  
A rough estimate of the initial amplitude of the transverse motions
within the shell can therefore be derived from the size of this cluster.  
A star located at a distance $r$ from the shell centre will induce a 
transverse velocity $v$ such that:

\begin{equation}
v ~\simeq ~V_s ~ \frac{r}{R_s}\;.
\vspace{11pt}
\label{eq:vt_1}
\end{equation} 

In order to estimate the size $r$ of the cluster of massive stars hosted by
a PGCC, 
we refer to R\,136, the dense core of the 30\,Doradus Nebula located in the
Large Magellanic Cloud.  The 30\,Doradus nebula shows an impressive example
of a two-stage stellar formation.  The energetic activity of a very compact 
bright cluster, R136, which includes several tens of O stars, triggers 
the formation of a new stellar generation revealed by numerous infrared 
sources in or near some bright filaments west and northeast of R136 
(Rubio et al.~1998).  Based on Hubble Space Telescope photometry, Campbell 
et al.~(1992) detected about 160 stars more massive than 10\,M$_{\odot}$ in R\,136
which they define as a region of 2.2\,pc $\times$ 1.9\,pc.  This number
of massive stars being remarkably similar to the numbers of SNeII used
in our self-enrichment model, we adopt R\,136 as the most similar example 
in the Local Group of what may have been the first generation cluster. 
A radius of 1\,pc appears therefore as a reasonable estimate of the size 
of this cluster, the source of the energy input.  \\

Considering an average background pressure of 10$^{-10}$\,dyne.cm$^{-2}$ 
and the corresponding cloud radius ($\simeq$ 30\,pc, see Eq.~3, Paper I), 
the transverse velocity when the shell reaches the cloud boundary is: 
\begin{equation}
v(t_{em}) ~\simeq ~\frac{1\,pc}{30\,pc} ~V_s(t_{em}) ~\simeq 0.03 V_s(t_{em})\;.
\label{eq:vt_2}
\end{equation}

Figure 5 displays the evolution with time of $\sigma _0$
and $\tilde \sigma _1$ assuming three different values for the initial
transverse velocity, namely 0.01\,$V_s(t_{em})$, 0.02\,$V_s(t_{em})$ and
0.03\,$V_s(t_{em})$, while keeping all the other parameters to their previous
values.  In sharp contrast with $\tilde \sigma _1(t_{em})$,
$\tilde v(t_{em})$ appears to be an important parameter of the shell
transverse collapse.  For instance, considering the bottom panel
in Fig.~5 ($P_h=5 \times 10^{-10}$\,dyne.cm$^{-2}$
and $N$=200), the fragmentation takes place less than 15 million years
after the first SN explosion if $v(t_{em})=0.03 \,V_s(t_{em})$, 
whereas it is prevented if $v(t_{em})=0.01 \,V_s(t_{em})$.    

\begin{figure}
\epsfig{figure=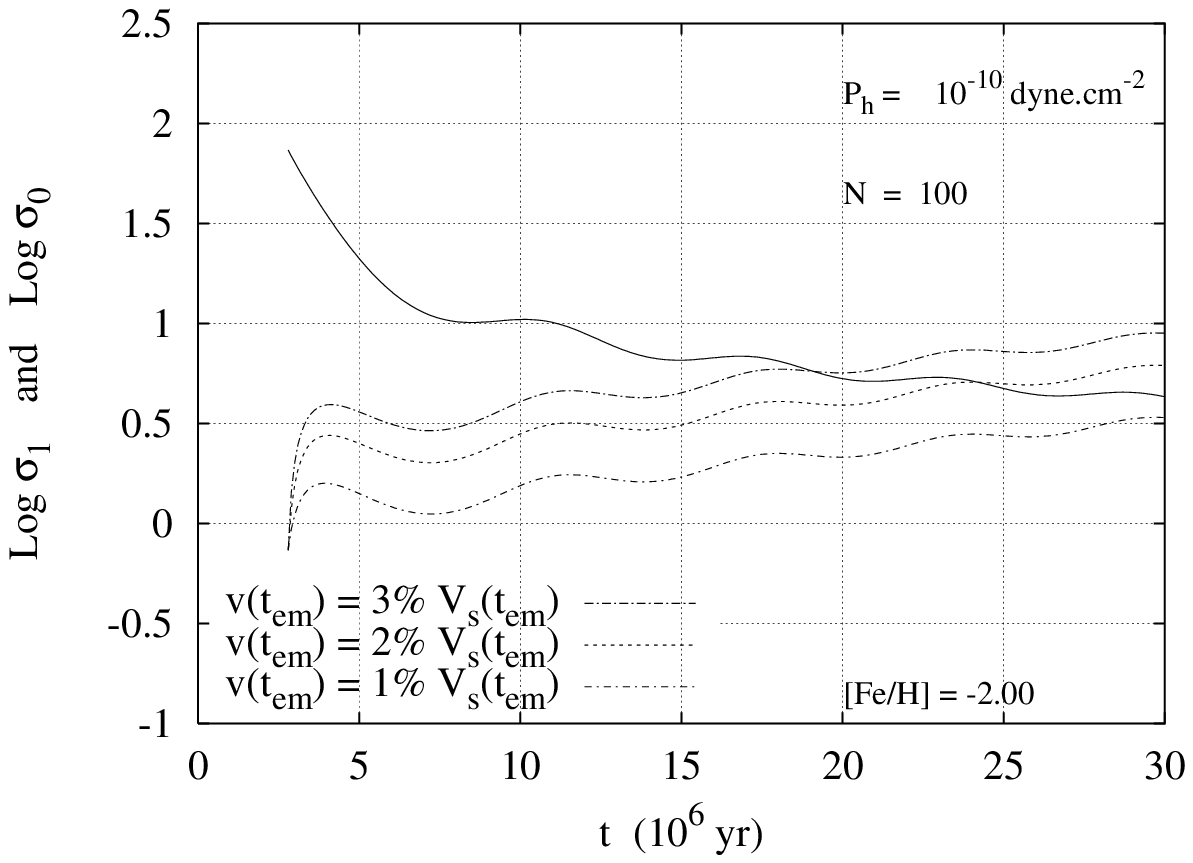, width=\linewidth}
\epsfig{figure=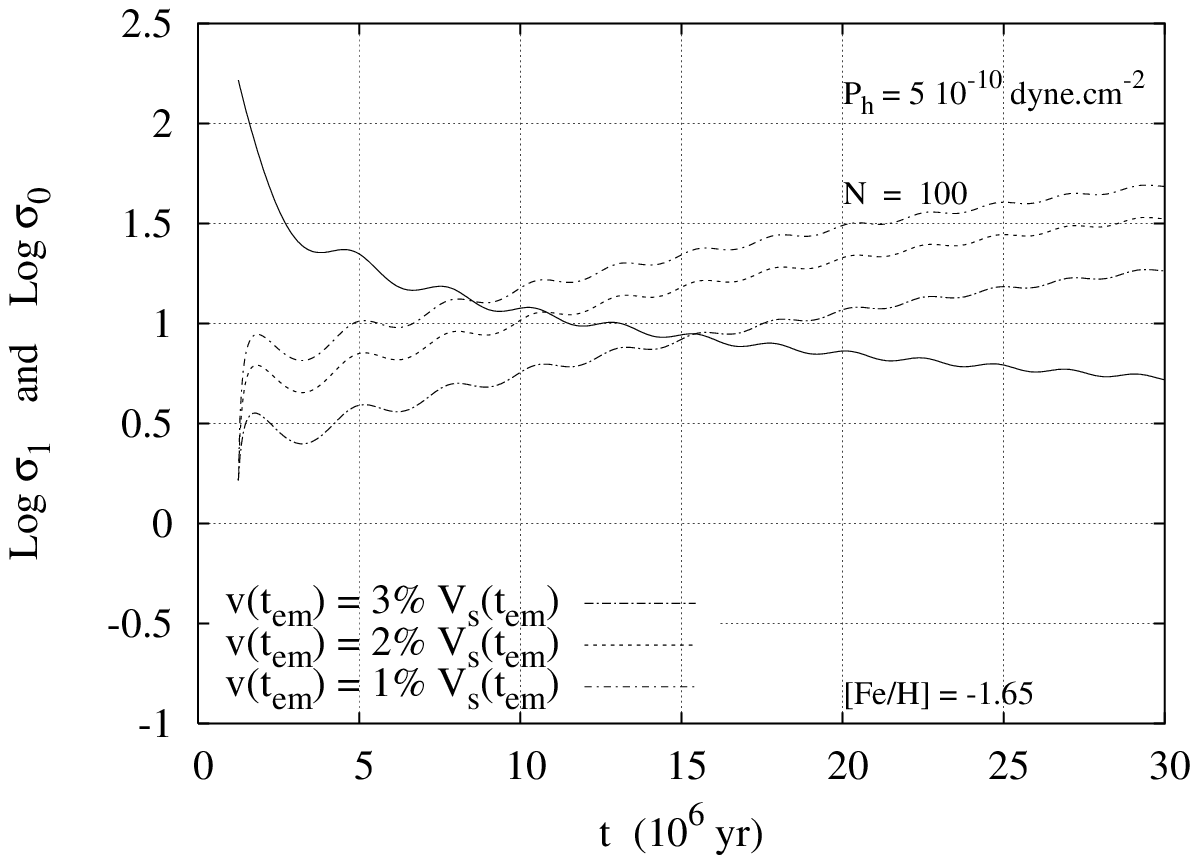, width=\linewidth}
\epsfig{figure=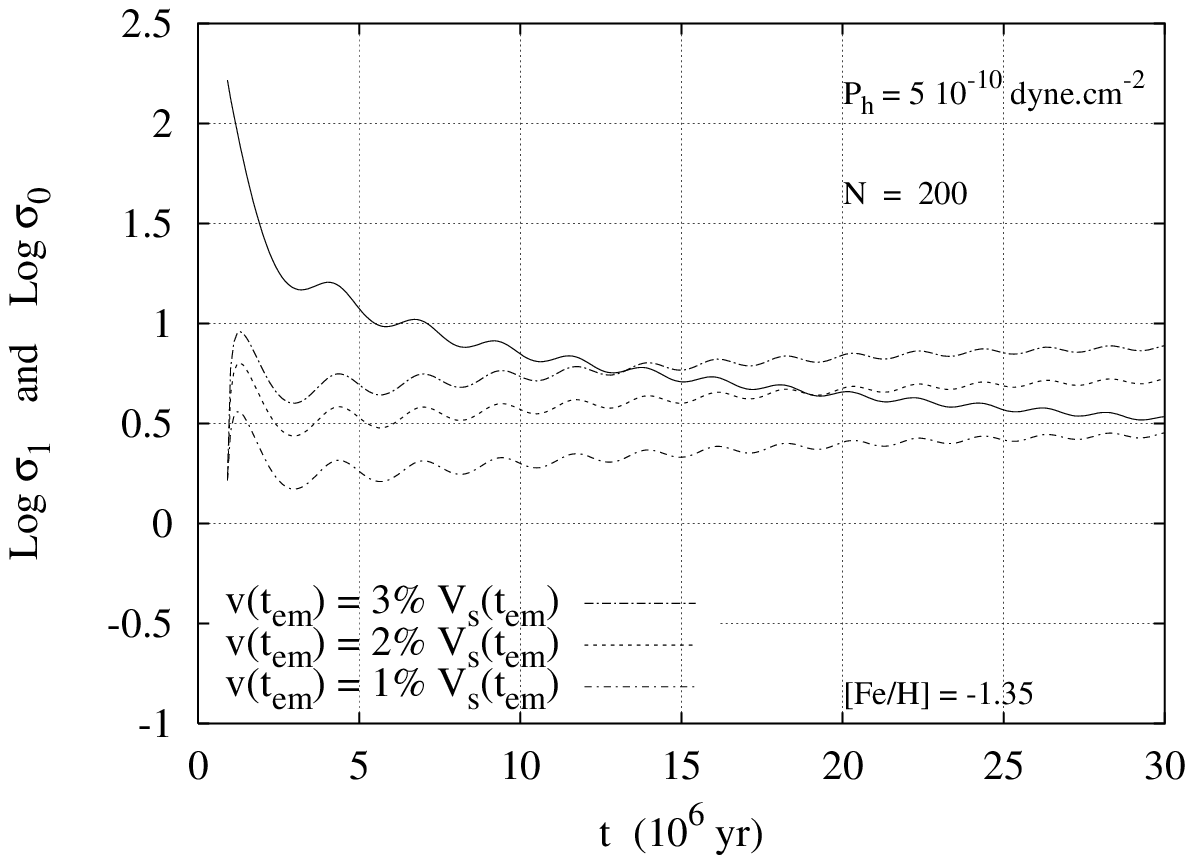, width=\linewidth}
\caption{Temporal evolution of $\sigma _0$ (plain curves) and 
$\tilde \sigma _1$ (dashed-dotted curves), expressed in units of 
1\,M$_{\odot}$.pc$^{-2}$, considering three initial perturbed velocities 
(see the keys) and the external pressures and SNII numbers
indicated.  In each panel, $c_s$=1\,km.s$^{-1}$, $\eta$=10 and  
$\tilde \sigma _1(t_{em})=0.01\,\sigma _0 (t_{em})$}
\vspace{11pt}
\vspace{11pt}
\label{fig:disc_vtrans}
\end{figure} 

\subsubsection{The number of forming clumps: $\eta$}
\label{sub:disc_eta}

\begin{figure}
\epsfig{figure=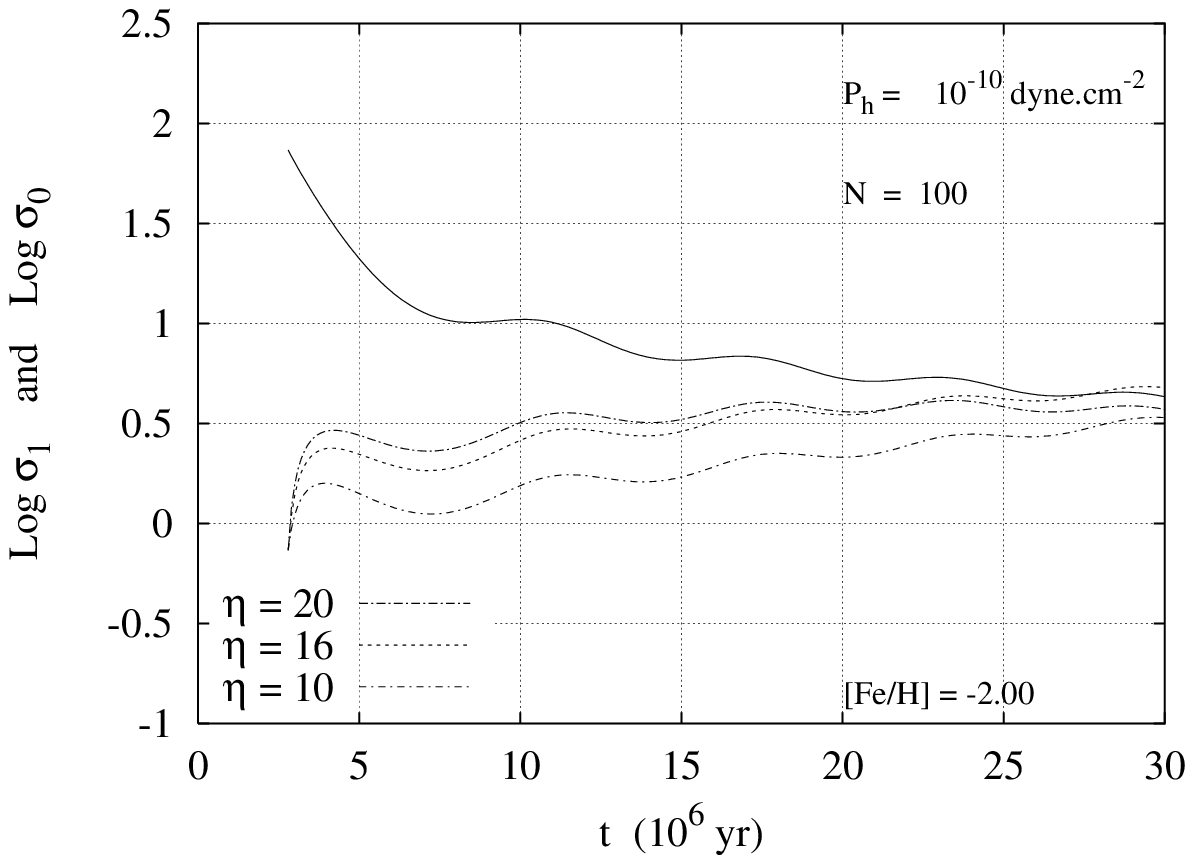, width=\linewidth}
\epsfig{figure=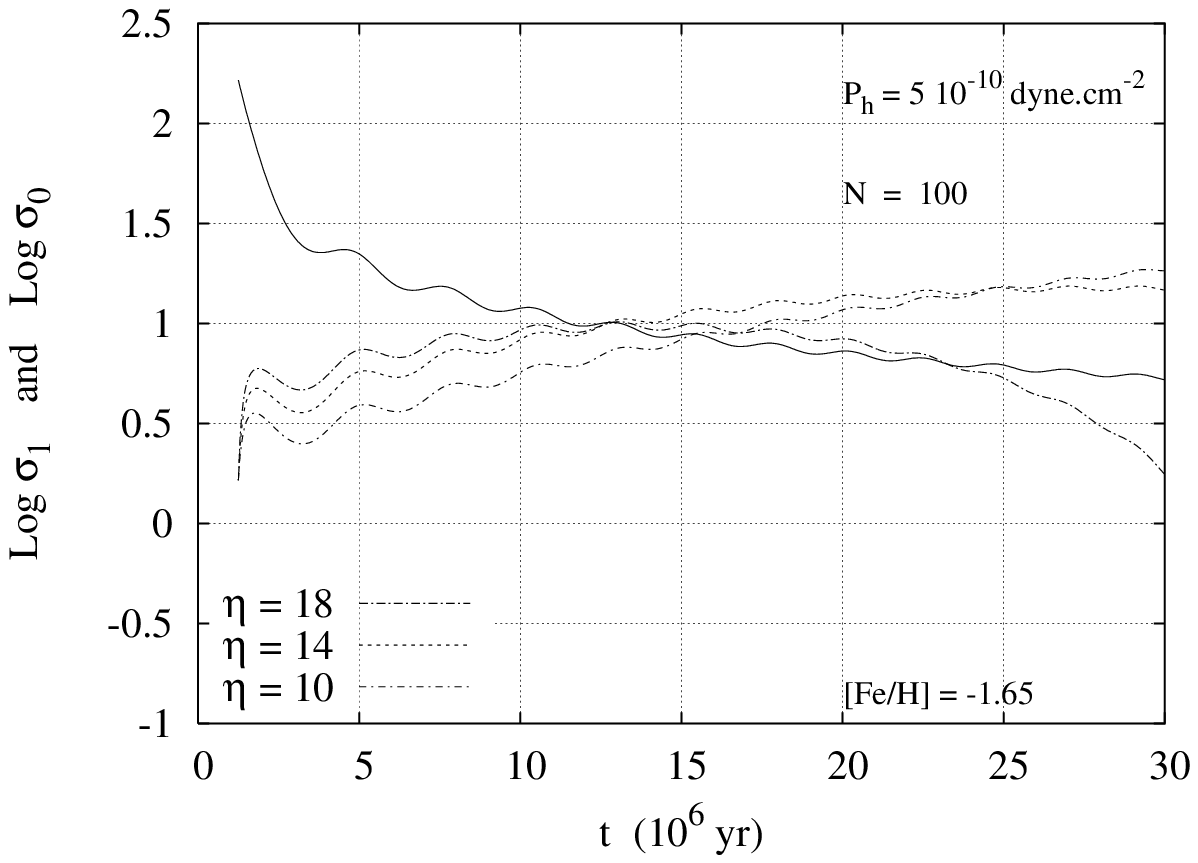, width=\linewidth}
\epsfig{figure=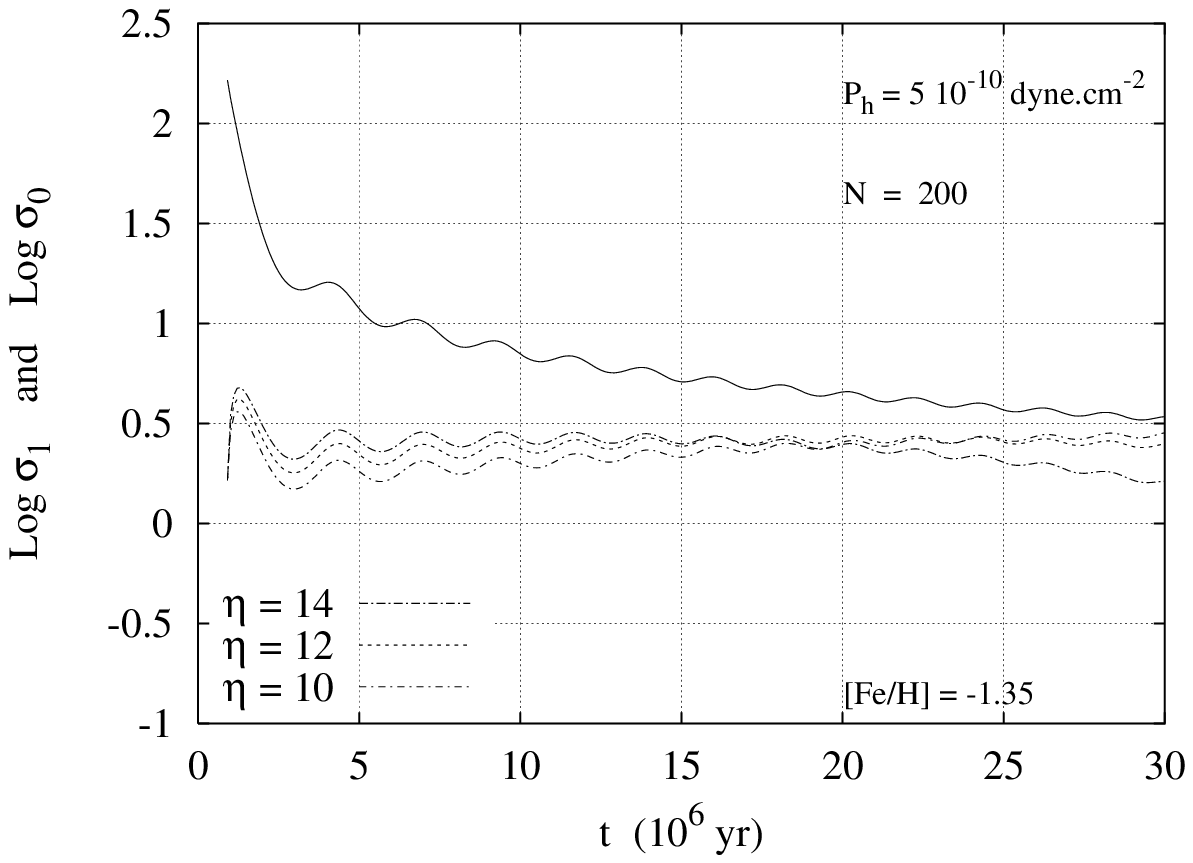, width=\linewidth}
\caption{Temporal evolution of $\sigma _0$ (plain curves) and 
$\tilde \sigma _1$ (dashed and dotted curves), 
expressed in units of 1\,M$_{\odot}$.pc$^{-2}$, considering
the external pressures, SNII numbers and numbers of forming clumps 
indicated (see the keys).  In each panel, $c_s$=1\,km.s$^{-1}$, 
$\tilde \sigma _1(t_{em})=0.01\sigma _0(t_{em})$, 
$v(t_{em})=0.01\,V_s(t_{em})$}
\vspace{11pt}
\label{fig:disc_eta}
\end{figure} 

In supernova-driven shells of gas, star formation does not take place
all along the whole periphery but, instead, takes place in regularly
spaced  clumps (i.e., where the transverse collapse brings some gas from 
depleted adjoining regions in the shell, see Fig.~\ref{fig:fragshell}).  
The number $\eta$ of such clumps along a shell circumference is thus 
related to the wavelength $\lambda$ of the
perturbation (Eq.~\ref{eta_lambda}).  This aspect of shell collapse
modelling following which stars are formed in discrete stellar subsystems
along a shell periphery is supported by several observations of star 
forming shells, the so-called Sextant being one of the most illustrative.  
In the Large Magellanic Cloud, a set of five OB associations are located 
along a HI supershell, sustaining there about 1/6 of a complete circle
and being thus called the Sextant (Efremov \& Elmegreen 1998).
Efremov, Ehlerova \& Palous (1999) showed that the OB associations are
regularly spaced, the deprojected average distance between two subsequent 
stellar groups being $\sim$ 37~pc.  Furthermore, using numerical simulations
of shells expanding in a mass model of the Large Magellanic Cloud, they
concluded that the formation of these 5 OB associations is most probably 
the result of a triggered star formation episode in the supershell created 
by SNeII located near the Sextant centre.   
Their simulations also explain how projection effects make the visible star 
forming regions only a fraction of a total circle.   \\
     
The influence of the number of clumps embedded within the shell
on the transverse collapse is illustrated in Fig.~\ref{fig:disc_eta}.
The parameter $\eta$ is not as negligeable as the initial perturbed 
surface density.  In some cases, it acts upon the fragmentation process 
almost as strongly as the initial transverse velocity. 
It is thus worth estimating the angular wavenumbers which are the most 
favourable to the growth of an initial perturbation.  
In order to do so, we now derive an analytical approximation of the 
instantaneous growth rate of the perturbation.  This is done straightforwardly
by assuming that the perturbed quantities vary exponentially with time $t$, 
in analogy with the exponential growth rate found in other instability 
problems (e.g. the Jeans mass) even though there is no real exponential 
growth in this problem owing to the shell expansion and the corresponding 
time dependence of $\sigma _0$.  It is thus important to keep in mind
that the perturbation growth rate derived below (Eqs.~\ref{eq:omega_eta}
and \ref{eq:fg_omega}) is {\it not} used to compute the temporal evolution 
of $\sigma _1$ (the cases displayed in Fig.~\ref{fig:disc_eta} are
obtained from solving 
Eqs~\ref{eq:per_cont_num_phi}-\ref{eq:per_motion_num_phi}) 
but merely to derive $\eta$ estimates favouring the shell collapse. 
The perturbed quantities are written:
\begin{equation}
\sigma _1 (t,\phi) = \tilde{\sigma}_1(t_{em}) ~e^{\omega t} ~e^{-i \eta \phi}
\label{sigma1_exp}
\end{equation}
and
\begin{equation}
v (t,\phi) = 
\tilde{v}(t_{em}) ~e^{\omega t}  ~e^{-i \eta \phi} ~e^{i \Delta \phi},
\label{v_exp}
\end{equation}

where $\omega$ is the angular frequency of the perturbation. 
Following Eqs.~\ref{sigma1_exp} and \ref{v_exp}, we write $\omega$ for the 
time derivatives and $- i \eta/R_s$ for the transverse gradients.
Therefore, Eqs.~\ref{eq:per_cont} and \ref{eq:per_motion}
become respectively

\begin{equation}
\omega \sigma _1 = -2 \frac{V_s}{R_s} \sigma _1
+ \sigma _0 \frac{i \eta}{R_s} v
\label{eq:per_cont2}
\end{equation} 

and

\begin{equation}
\sigma _0 \omega v = - \sigma _0 \frac{V_s}{R_s} v
+ {c_s}^2~ \frac{i \eta}{R_s}\sigma _1 - 2 \pi i G \sigma _0 \sigma _1 \,, 
\label{eq:per_motion2}
\end{equation}  

using Eq.~\ref{g1_sigma1} in the latter.  \\

The elimination of the perturbed quantities $\sigma _1$ and $v$  
between Eqs.~\ref{eq:per_cont2} and \ref{eq:per_motion2} provides 
the dispersion equation, namely the relation between 
the angular frequency $\omega$ (i.e. the instantaneous growth rate) 
and the angular wavenumber $\eta$ of the perturbation:
\begin{equation}
\sigma _0 \left( \omega + 2 \frac{V_s}{R_s} \right)
\left( \omega + \frac{V_s}{R_s} \right) - \sigma _0 \frac{i \eta}{R_s}
\left( 2 \pi i G \sigma _0 - c_s^2 \frac{i \eta}{R_s} \right) = 0\,,
\end{equation}
whose solution is given by 
\begin{equation}
\omega (\eta) = -\frac{3}{2} \frac{V_s}{R_s} + 
\sqrt{\frac{3}{2} \frac{V_s ^2}{R_s ^2} 
+ 2 \pi G \sigma _0 \frac{\eta}{R_s} - c_s^2 \frac{\eta ^2}{R_s ^2}}.
\label{eq:omega_eta}
\end{equation}

Let us consider the {\sl first growing mode}.  This one corresponds to the 
sequence of values of $\eta$ which maximises the angular frequency $\omega$ 
at each moment of the shell propagation, i.e. $\eta _{fg} = \eta (t)$ such that
\begin{equation}
\frac{d \omega}{d \eta} = 0\;.
\label{eq:fg_def}
\end{equation}   
Equation \ref{eq:fg_def} indeed corresponds to a maximum since the 
discriminant of Eq.~\ref{eq:omega_eta} shows a negative curvature with $\eta$. 
The instantaneous angular wavenumber and the instantaneous angular
frequency associated to the first growing mode obey respectively
\begin{equation}
\eta _{fg} = \frac{\pi G}{{c_s}^2} \sigma _0 R_s 
= \frac{1}{4 {c_s}^2} \frac{GM}{R_s}
\label{eq:fg_eta}
\end{equation}  

and

\begin{equation}
\omega _{fg} = - \frac{3}{2} \frac{V_s}{R_s} 
+ \sqrt{\frac{V_s^2}{R_s^2} + \frac{\pi ^2 G^2 \sigma _0^2 }{c_s^2} }\;.
\label{eq:fg_omega}
\end{equation}  

Taking into account the dependence of $M$ and $R_s$
(Eq.~\ref{eq:RsHPBt1/3}) on $P_h$ and $N$, Eq.~\ref{eq:fg_eta} shows that
$\eta _{fg}$ depends on the external pressure $P_h$, on the SN number $N$
and on time $t$ as
\begin{equation}
\eta _{fg} \propto P_h ^{-1/6} N^{-1/3} t^{-1/3}\;.
\label{eq:eta_fg/NPh}
\end{equation}

A first guess of a favourable angular wavenumber can be estimated from
the temporal average of Eq.~\ref{eq:fg_eta} over the time
spent by the supershell in the hot protogalactic background:
\begin{equation}
<\eta _{fg}> = \frac{1}{\Delta t - t_{em}} \int_{t_{em}}^{\Delta t}
\eta _{fg}(t') ~ dt' \,,
\label{eta_fg_ave}
\end{equation}

where $\Delta t$ is the duration of the SN phase.  Among the cases
displayed in Fig.~\ref{fig:disc_eta} ($c_s$=1\,km.s$^{-1}$, 
$\tilde \sigma _1(t_{em})=0.01\sigma _0(t_{em})$, 
$v(t_{em})=0.01\,V_s(t_{em})$), we see that the shell transverse 
collapse is achieved if, for instance, $P_h = 10^{-10}$\,dyne.cm$^{-2}$,
$N$=100 and $\eta$=16 (top panel) or if 
$P_h = 5 \times 10^{-10}$\,dyne.cm$^{-2}$, $N$=100 and $\eta$=10
(middle panel).  It is interesting to note that these values
of $\eta$ reasonably match those given by Eq.~\ref{eta_fg_ave},
for the above mentioned cases, i.e., $<\eta _{fg}>$=12 and 
$<\eta _{fg}>$=9, respectively.  As Eq.~\ref{eta_fg_ave}
has been derived under the assumption of an exponential growth rate, 
this agreement a posteriori justifies its validity as a convenient 
estimate of the number of forming clumps in collapsing shells.  \\

\subsubsection{The shell sound speed: $c_s$}
\label{sub:disc_cs}

\begin{figure}
\begin{center}
\epsfig{figure=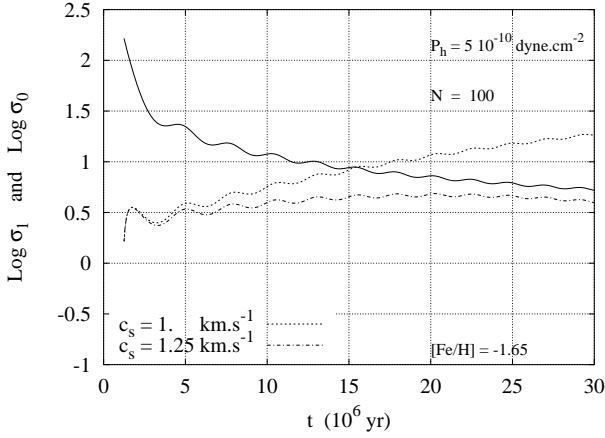, width=\linewidth}
\label{fig:disc_cs}
\caption{Temporal evolution of $\sigma _0$ (plain curve) and 
$\tilde \sigma _1$ (dashed curves), expressed in units of 
1\,M$_{\odot}$.pc$^{-2}$, considering the sound speeds (see the key),
 external pressures and SNII numbers indicated.  Other parameters are 
$\eta$=10,  $\tilde \sigma _1(t_{em})=0.01\sigma _0(t_{em})$, 
$v(t_{em})=0.01\,V_s(t_{em})$}  
\end{center}
\end{figure}

The development of a gravitational instability within a supershell also
depends on the sound speed $c_s$ of the shell material.
Indeed, $c_s$ is directly related to the thermal pressure $P_s$ of the shell
gas through 
\begin{equation}
P_s = \frac{\rho _s k T_s}{\mu _s m_H} = \rho _s c_s ^2\,.
\end{equation}
In this equation, $k$ and $m_H$ are the Boltzmann constant and the hydrogen 
mass, respectively, while $\rho _s$, $T_s$ and $\mu _s$ are the mass density, 
the temperature and the mean molecular weight of the shell, respectively.  
The larger the velocity dispersion, the more the layer resists 
gravitational collapse.
In an H\,I layer (T $\simeq$ 100\,K, $\mu ~\simeq$ 1.3\,m$_H$), the sound 
speed is $\simeq$ 0.8\,km.s$^{-1}$, while it can be as low as 
$\simeq$ 0.3\,km.s$^{-1}$ in an H$_2$ layer (T $\simeq$ 20\,K, 
$\mu ~\simeq$ 2.1\,m$_H$) (McCray \& Kafatos 1987).  However,
the turbulence and magnetic fields of the shell will increase these values.
To some extent, they can be represented by an additional pressure term 
in the expression of $c_s$, i.e.
\begin{equation}
c_s = \left ( \frac{k T_s}{\mu _s m_H} 
+ \frac{B_s ^2}{4 \pi \rho _s} + \mathcal{T} {\rm _s} \right)^{1/2}  
\label{eq:cs_mag}
\end{equation} 
where $B_s$ is the magnetic field in the shell and $\mathcal{T} {\rm _s}$ 
is the contribution of turbulence.   \\

Figure 7 shows how highly sensible to $c_s$
the fragmentation process is.  
Considering 100\,SNeII and a hot background pressure of
5 $\times$ 10$^{-10}$\,dyne.cm$^{-2}$, the fragmentation takes place
even with a low initial transverse velocity, i.e. 
$v(t_{em})=0.01 V_s(t_{em})$, if $c_s$=1\,km.s$^{-1}$.  Increasing the latter
by 25\%, the transverse collapse is very weakened and the
fragmentation is prevented.  During the last ten million years, the
evolutions with time of $\sigma _0$ and $\tilde \sigma _1$ are similar, 
that is, the evolution of $\tilde \sigma _1 (t)$ is mostly driven by 
the dilution of $\sigma _0 (t)$ due to the shell expansion. \\
 
Equation \ref{eq:cs_mag} illustrates the difficulty of
estimating the sound speed of a gas.  It implies the computations
of its cooling history, its magnetic fields and turbulence.
Moreover, the high sensibility of the fragmentation issue
to $c_s$ (see Fig.~7) shows that its value must 
be estimated with some accuracy.  Such a task is well beyond the scope of 
the present work and, in what follows, we adopt $c_s=1\,{\rm km.s}^{-1}$, 
in agreement with many studies of supershell fragmentation (e.g. 
Comeron \& Torra 1994, Ehlerova \& Palous 2002).

\section{Discussion}
\label{sec:search_para}

\subsection{Stars with halo GC metallicities} 
\label{sec:disc_GC}

The previous section has shown that the shell/swept PGCC may become 
gravitationally unstable and finally break into fragments providing
that some conditions are fulfilled, e.g. 
$v(t_{em})/V_s(t_{em})  \simeq $ 0.03, $c_s \simeq $ 1\,km.s$^{-1}$, 
$\eta \simeq <\eta _{fg}>$.  This is {\sl not} to claim that all 
supershells will encounter such favourable circumstances, but one 
may expect that at least {\sl some} of them will do. \\

\begin{figure}
\begin{minipage}[b]{\linewidth}
\begin{center}
\epsfig{figure=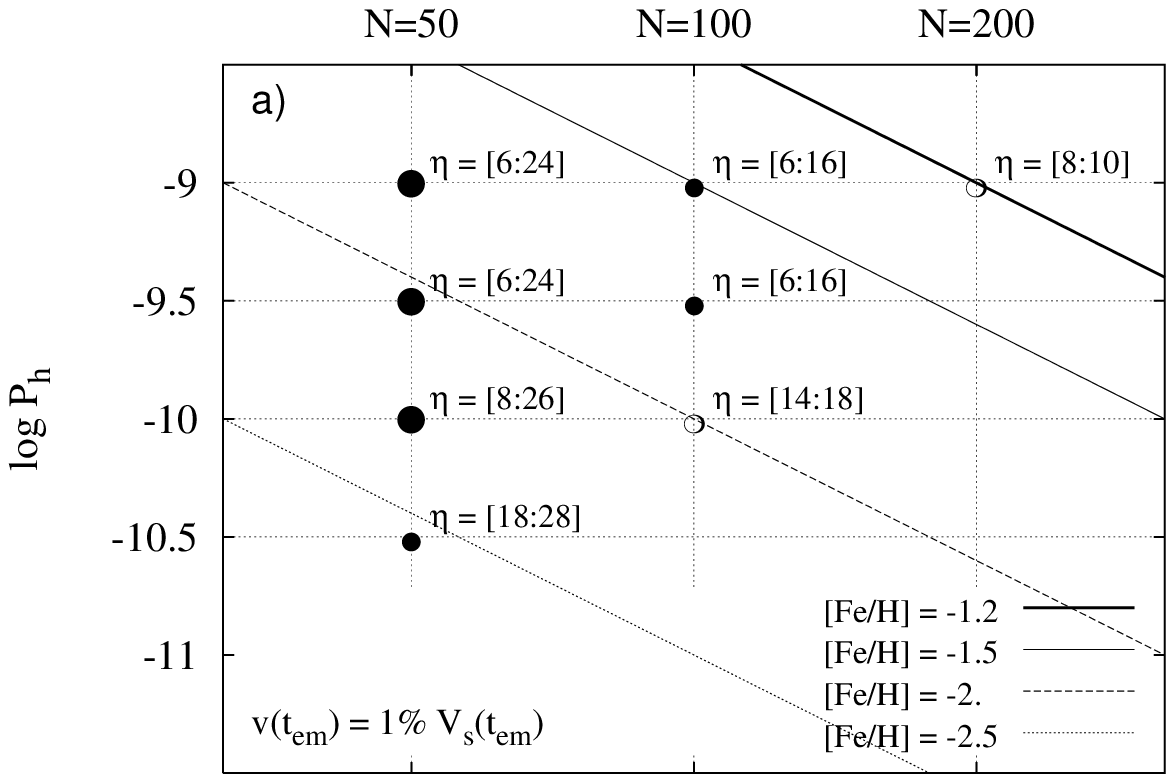, width=\linewidth}
\end{center}
\end{minipage}
\vfill
\vspace*{-9mm}
\begin{minipage}[b]{\linewidth}
\begin{center}
\epsfig{figure=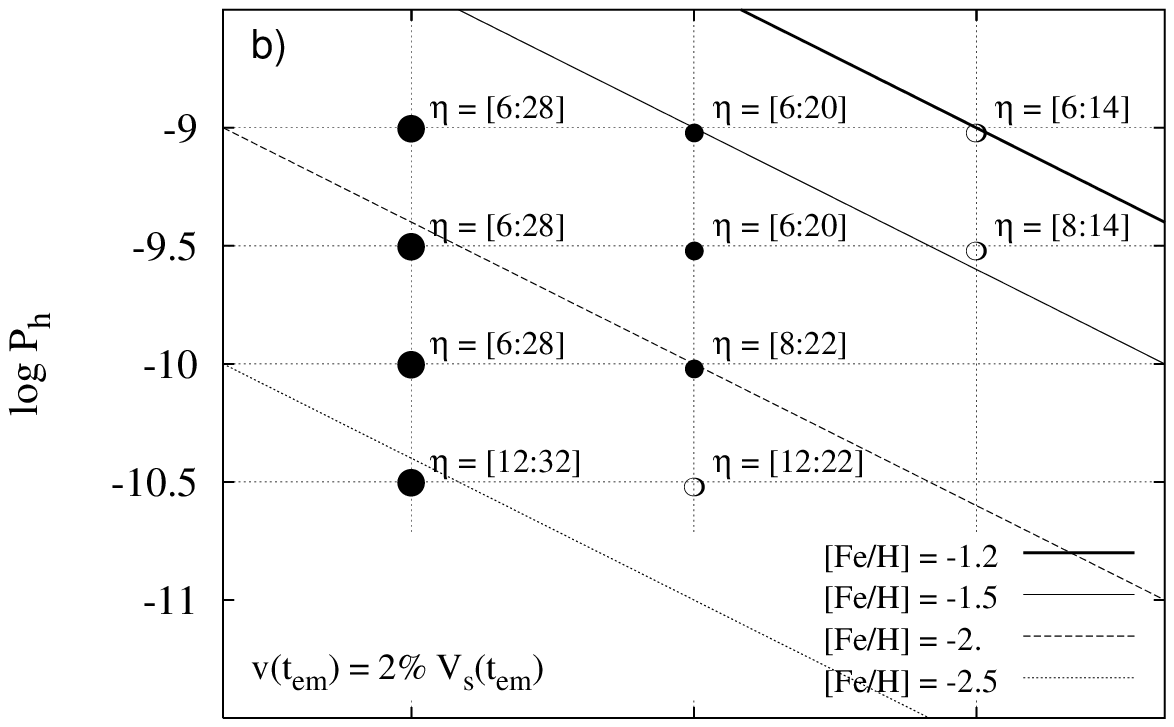, width=\linewidth}
\end{center}
\end{minipage}
\vfill
\vspace*{-9mm}
\begin{minipage}[b]{\linewidth}
\begin{center}
\epsfig{figure=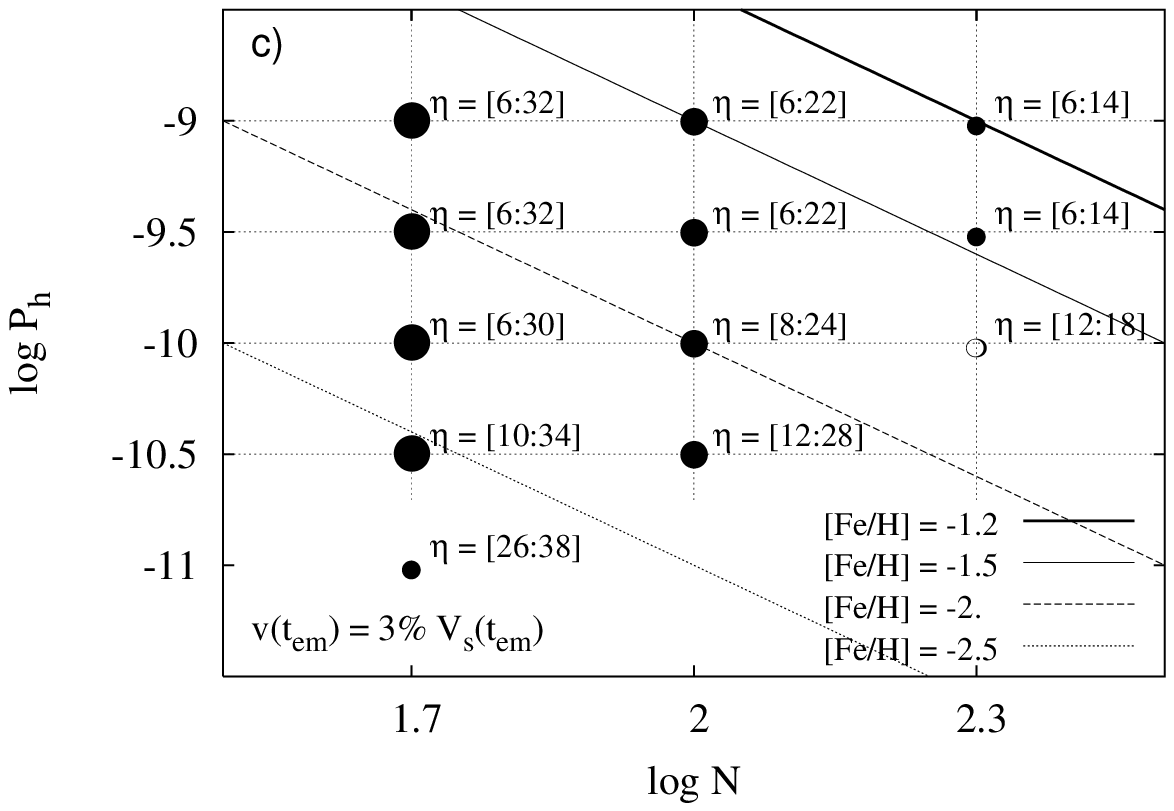, width=\linewidth}
\end{center}
\end{minipage}
\label{fig:frag_prob}
\caption{Summary of the ($N$, $P_h$, $v(t_{em})/V_s(t_{em})$, $\eta$) values 
which may (open or filled circles), 
or may not (no symbol), lead to a swept PGCC fragmentation  
providing that the transverse flows converge towards the clumps embedded 
within the shell, i.e. $\Delta \phi = -\pi /2$.  The sound speed 
and the initial perturbed surface density are $c_s = 1$\,km.s$^{-1}$ and 
$\sigma _1 (t_{em}) = 0.01 ~\sigma _0 (t_{em})$, respectively.
The top, intermediate and bottom panels correspond to initial
transverse velocity of 1, 2 and 3 per cent of the shell 
velocity when it emerges out of the cloud, respectively.  The 
isometallicity curves corresponding to [Fe/H]=$-$1.2, $-$1.5, $-$2, $-$2.5 
are also plotted, giving thus the metallicity achieved through 
self-enrichment for each parameter combination.  For each ($N$,$P_h$) pair, 
54 couples of initial perturbed velocities ($v(t_{em})/V_s(t_{em}$)=
0.01, 0.02, 0.03) and clump numbers ($\eta$ ranging from 6 to 40 by step 
of 2) are tested.    In each panel (i.e., for each value of the initial 
perturbed velocity), a couple ($N$, $P_h$) leading to a successful 
transverse collapse is marked by an open/filled circle as well as by 
the range of $\eta$ values leading to fragmentation.  The larger the range 
of $\eta$ values, the bigger the symbol (see text for details)}
\end{figure} 

Figure 8 presents the results of shell fragmentation
simulations for varying values of $N$, $P_h$, $\eta$ and 
$v(t_{em})/V_s(t_{em})$, assuming that $c_s$=1\,km.s$^{-1}$ and 
$\sigma _1 (t_{em}) $= 0.01 $\sigma _0(t_{em})$. Five hot protogalactic 
background pressures ($P_h=10^{-11}, 3.2 \times 10^{-11}, 10^{-10},
3.2 \times 10^{-10}, 10^{-9}$\,dyne.cm$^{-2}$) and 3 SN numbers ($N$=50, 
100, 200) are tested.  The upper limit for 
$N$ is the maximum number of supernovae that the GC gaseous progenitor
can sustain, namely $N$=200 (i.e., if $N > 200$, the absolute value 
of the cloud binding energy is lower than the shell kinetic energy: 
disruption criterion, Paper I).  The lower 
limit is imposed by $c^{PGCC}$, the sound speed of
the PGCC material.  Indeed, the shell is ``built'' while sweeping the
PGCC and such a mass accumulation into a shell requires the velocity 
of the shell to be larger than the sound speed of the ambiant medium.  
Therefore, the lower limit to the shell velocity in the PGCC obeys:
\begin{equation}
{V_s}^{PGCC} = {c}^{PGCC} = \sqrt{\frac{k T}{\mu m_H}} 
= 8.3\,{\rm km.s^{-1}}  
\end{equation}
where $T$ and $\mu$ are the temperature ($\simeq 10^4$\,K) and the mean 
molocular weight ($\simeq 1.2$) of the PGCC, respectively (Fall \& Rees 1985).
The number of SNeII corresponding to this lower limit of the shell velocity  
is $N \simeq 50$ (Eq.~13, Paper I).
Regarding the upper value of $P_h$, we refer to Murray \& Lin (1992)
who showed that the hot protogalactic background pressure depends on
the galactocentric distance $D$ as
\begin{equation}
P_{h} = 1.25\times 10^{-9} D_{kpc}^{-2}  \,\, \rm dyne\,cm^{-2}.
\label{Ph2}
\end{equation}
Thus, $P_h$ is of order 10$^{-9}$\,dyne.cm$^{-2}$ in the very inner 
Galactic regions (i.e., $D \simeq$ 1\,kpc).  \\
In order to see what is the metallicity achieved in the shells which 
succeed in forming new stars, iso-metallicity curves corresponding to 
[Fe/H]=$-$1.2, $-$1.5, $-$2 and $-$2.5 (i.e., metallicities typical of 
the Galactic halo GCs) are displayed in each panel of Fig.~8. \\    
For each couple ($N$,$P_h$), 54 combinations of $v(t_{em})/V_s(t_{em})$ 
and $\eta$ (i.e. 3 initial perturbed velocities $\times$ 
18 numbers of clumps) have been run.  The initial transverse velocities 
correspond to 1, 2 and 3 per cent of the velocity of the shell when it 
enters the hot background and the number of clumps ranges from 6 to 40
by step of two.  It appears that the shell is unable to fragment if 
$\eta \ge 40$.  When a set of conditions leads to a successful 
fragmentation, the corresponding point in the ($N$, $P_h$) diagram is 
marked by a circle  as well as by the range of $\eta$ values leading 
to successful transverse collapses.  Depending on whether
the shell fragmentation takes place for 0 to 25 per cent, 25 to 50 per
cent, 50 to 75 per cent, or 75 to 100 per cent of the number of $\eta$ 
values tested, the corresponding triplet 
($N$, $P_h$, $v(t_{em})/V_s(t_{em}$)) is marked by an open circle, a small, 
a median-size or a large filled circle, respectively.    
Figure 8 confirms that the probability of successful transverse collapse 
fades away with decreasing external pressure and increasing number of SNeII.  
As mentioned in Sect.~\ref{sub:discNPh}, a low external pressure and a 
high number of SNeII leads to a larger radius for the shell, decreasing 
thereby its surface density (Eq.~\ref{eq:sig0_N_Ph}) and its ability 
to collapse.   \\

At high pressure, i.e. $P_h \simeq 10^{-9}$\,dyne.cm$^{-2}$, the ability 
of the shell to get fragmented is limited by too large a number of SNeII.  
Figure 8 shows however that a SN number as large as 200 does not prevent the 
fragmentation.  Accordingly, the largest metallicity which can be 
achieved through self-enrichment is [Fe/H]$\simeq -$1.2.
On the other hand, at low pressure, the transverse collapse is supported 
by a low number of SNeII.  Combining a low background pressure with the 
lower limit on $N$, Fig.~8 shows that the lowest 
metallicity which can be achieved is [Fe/H]$\simeq -$2.8 providing that
the relative initial transverse velocity is 3 per cent.  
While this lower limit in metallicity is a bit uncertain, as it is achieved
in the case of the highest intial transverse velocity only, the examination
of the three panels in Fig.~8 clearly shows that
a metallicity of $\simeq -2.5$ is actually achievable.  These extreme 
values ([Fe/H]$\simeq -$1.2 and $-$2.5) match nicely the metallicities 
exhibited by the most metal-rich and the most metal-poor Galactic halo 
GCs, respectively. \\

At this stage, it is worth keeping in mind that the question addressed here
above concerns the ability of the shell to form stars and not yet the 
ability of these stars to evolve into a stellar cluster.
We will address the ability of the newly 
formed stars to form a bound cluster in a forthcoming paper and we  
emphasize here that, among the cases of successful fragmentation displayed by 
Fig.~8, some of the newly formed stars may not be able 
to form a bound cluster.  In other words,  the collapsed shells may be a 
source for both halo GCs and halo field stars.  

Section 2.3.2 has discussed how the self-enrichment model for
GC formation can shed light on the metallicity distribution of the
Galactic halo.  Because the parameters which determine the metallicity, 
i.e. $N$ and $P_h$, also control the shell surface density, there
is a direct link between the metallicity and the probability of star 
formation\footnote{This is {\sl not} to say that the probability
of star formation is determined by the metallicity, as it can be
assumed in pre-enrichment models in which the metallicity may control the
star formation through line cooling processes.  It rather means 
that the metallicity and the probability of star formation derive from the 
same parameters, namely the number of SNeII and the pressure of the hot 
protogalactic background}.  Furthermore, the initial radius and velocity 
of the second generation stars 
being the radius and the velocity of the shell at the time of their 
formation, their binding depends on the shell expansion law and, 
therefore, again, on $N$ and $P_h$.  As a consequence, the probability of 
getting a bound cluster from a shell of newly formed stars is also related 
to their metallicity.  If we assume that a significant fraction of the
field content comes from ``failed GCs'' (i.e. shells of stars which did not 
succeed in forming GCs), then these relations between 
the metallicity on the one hand and the probabilities of forming a 
second stellar generation and a bound cluster on the other hand  
{\sl open the way to the computation of the metallicity distributions
of both halo field stars and halo GCs}. 

Since the formation of a GC requires one more condition than the shell 
transverse collapse (i.e. the binding of the second stellar generation),
only a subset of the ($N$,$P_h$) combinations giving rise to 
shell fragmentation may lead to the formation of a bound cluster.  Such a 
conclusion does not contradict, and even fits, the fact that the metallicity
spectrum of halo field stars is larger than the one of GCs
(e.g. Laird et al.~1988).  \\

\subsection{The very metal-poor stars ([Fe/H] $\lesssim -2.5$)}
\label{sub:vmp_st}

The most striking difference between the halo GC and halo field metallicity
distributions resides in the much more extended metal-poor tail of the field
compared to the GC system.  In fact, while the most metal-poor GCs show 
[Fe/H]$\simeq -$2.5, the Galactic halo hosts field stars whose 
metallicity is much lower (e.g., CS22876-032, [Fe/H]$\simeq -$3.7, Norris, 
Beers \& Ryan 2000).  The recent discovery of HE0107-5240, the most 
metal-deficient star ever discovered ([Fe/H]$\simeq -$5.3, Christlieb et 
al.~2002) has decreased even more the lower limit of the observed 
metallicity distribution of halo field stars.  Our model does not seem to 
be able to explain these very metal-poor stars (see Fig.~8)
and we must therefore investigate alternative scenarios for the formation
of these stars more metal-deficient than the most metal-poor GCs.

\subsubsection{Isolated Type II Supernovae}
The existence of very metal-poor stars has often been attributed
to star formation episodes triggered by {\it individual} SNeII exploding 
in primordial gas clouds (e.g., Shigeyama \& Tsujimoto 1998, Argast et 
al. 2000, Karlsson \& Gustafsson 2001), that is, a formation scenario 
similar to the self-enrichment scenario for GCs except regarding the 
considered number of massive stars.
The metallicity of such stars is determined by the ratio of the mass of 
metals ejected by the SNII to the mass of hydrogen gathered by the 
shock wave.  The SNII yields computed by Woosley \& Weaver (1995) for 
zero metallicity stars show that a SNII with progenitor mass $m$ ejects
a mass of metals $m_z$ such that $m_z \simeq 0.3m - 3.5$ (in unit of 
one solar mass).  On the other hand, the mass $M_{sw}$ of interstellar gas 
collected by the shock wave is related to the explosion energy $E_0$ through 
\begin{equation}
M_{sw} = 5.1 \times 10^4 M_{\odot} \left( \frac{E_0}{10^{51}\,ergs} 
\right)^{0.97}\,,
\label{single_SN}
\end{equation} 
assuming a sound speed of 10\,km.s$^{-1}$ (or T $\sim$ 10$^4$\,K) for the
interstellar gas (Shigeyama \& Tsujimoto 1998).  Using these relations 
and assuming $E_0 = 10^{51}$\,ergs for a canonical SNII, 
the metallicity achieved in shells of gas driven by isolated SNeII 
is straightforwardly derived (see the second column of Table 1).
The explosion energy being the same whatever the SNII progenitor mass,
every SNII sweeps the same amount of primordial circumstellar gas,
irrespective of the SNII mass (Eq.~\ref{single_SN}).  Therefore, the 
final metallicity is determined solely by the mass of heavy elements 
released in the interstellar medium by the exploding star and, thus, by 
the SNII progenitor mass.  As a consequence,
the more massive the SNII, the larger the final metallicity.
As already noticed by Shigeyama \& Tsujimoto (1998), the formation of very
metal-poor stars with [Fe/H] $\simeq -4$ and [Fe/H] $\simeq -2$ may
therefore be ascribed to ``low''-mass (i.e., M $\simeq$ 12\,M$_{\odot}$)
and high-mass (i.e., M $\simeq$ 40\,M$_{\odot}$) SNII, respectively.
In order to dig out the potential link between very metal-poor stars
and isolated SNeII, Shigeyama \& Tsujimoto (1998) also exploited
the yields of core-collapse SN predicted by nucleosynthesis 
calculations (e.g., Woosley \& Weaver 1995, Tsujimoto et al.~1995).  
They focused on the abundance patterns of C, Mg, Si and Ca.
The yields of these elements being not (or at least less) affected 
by the mass-cut issue, they should therefore be known with a better 
accuracy than the yields of heavier elements such as the iron peak ones.
They noticed the good agreement between the abundance patterns 
arising in the remnants of first generation SNeII, as predicted
by nucleosynthesis calculations, and those inferred from the spectra 
of halo stars with [Fe/H] $\lesssim -$2.5.  As a result, 
Shigeyama \& Tsujimoto (1998) suggested that very metal-poor stars 
actually formed out of interstellar gas which had been swept by 
isolated SNeII.  They also note that the large observational scatter in 
the abundance ratios among these stars can be explained by the differences 
in SNII yields with the mass of the progenitor.

It is worthy of a mention that such isolated SNeII are certainly not able
to trigger the formation of GCs owing to the small amount of interstellar 
gas swept by the blast wave (Eq.~\ref{single_SN}; also, not all the gas
will be converted into stars).  Should such low-mass halo clusters have
managed to form however, they will have been quickly destroyed by the 
Galactic tidal fields, low-mass clusters being among the most vulnerable
in this respect (e.g., Gnedin \& Ostriker 1997).  
Therefore, this extension of the self-enrichement scenario to single 
SNII explosions does not contradict the absence of Galactic 
halo GCs with [Fe/H] $\lesssim -$2.5. 

\subsubsection{Hypernovae}
In fact, even stars as massive as 40\,M$_{\odot}$ may trigger the formation
of stars with a metallicity as low as $-4$, providing that they explode
as {\it hypernovae}, i.e., supernovae characterized by explosion energies 
of order $E_0 \sim 10^{52} - 10^{53}$~ergs.  This class of objects includes 
two categories depending on the progenitor mass, namely core-collapse 
hypernovae and pair-instability hypernovae.
They both have been invoked as possible explanations to peculiar 
abundance patterns observed in some very metal-poor field stars.  \\

{\it Core-collapse hypernovae.  }  
Recent observations suggest that at least some core-collapse supernovae
explode with explosion energies ten to one hundred times higher than 
the energy released by a canonical SNII (e.g., Galama et al.~1998).  They
likely originate from relatively massive star (M$\gtrsim 25$~M$_{\odot}$). 
In contrast to pair-instability hypernovae (see below), the mass range of 
their progenitor is thus similar to the one of canonical SNeII.

If hypernovae occured in the early stage of the Galactic evolution and 
induced star formation, their abundance pattern may still be observable in 
the atmospheres of some low-mass halo stars.  
As for the case of star formation triggered by single SNeII, we can derive 
an estimate of the metallicity of these halo stars using 
Eq.~\ref{single_SN} and hypernova yields.  Nakamura et al. (2001) have
investigated in detail the nucleosynthesis of core-collapse hypernovae
and compared it with the yields of canonical (i.e., $E_0 \sim 10^{51}$\,ergs) 
core-collapse SNeII with similar progenitor mass.
Their Tables 2-5 show that SNeII and hypernovae with similar progenitor 
mass release much the same amount of metals.  The resulting metallicity 
of stars formed in hypernova remnants are given in 
Table 1, following the same of line of reasoning as in the previous section.  
Hypernovae being much more energetic than SNeII, and the mass of interstellar 
gas swept by the blast wave being roughly 
proportional to the explosion energy (Eq.~\ref{single_SN}), they will 
collect a larger amount of interstellar gas.  Therefore, should
core-collapse hypernovae be able to trigger the formation of new stars, 
these stars will be more metal-poor than the ones formed in the remnants 
of canonical SNeII.   Thus, they may be well-suited to explain the 
formation of stars with a metallicity as low as [Fe/H] $\simeq -4$ 
(see Table 1).

\begin{table}
\caption[]{Dependence of the metallicity [Fe/H] on the progenitor mass 
$m$ and explosion energy E$_0$ of supernovae 
(SN; E$_0 = 10^{51}$\,ergs) and hypernovae 
(HN; E$_0 = 10 ~~ {\rm or} ~~ 100 \times 10^{51}$\,ergs).  
Every mass is expressed in units of one solar mass.  
(1) A 12\,M$_{\odot}$ star explodes as a canonical SNII. According to 
Nakamura et al.~(2001), stars less massive than 25\,M$_{\odot}$ do not
explode as core-collapse hypernovae and such cases are thus not considered 
in this Table. (2-3) More massive stars explode either as 
core-collapse supernovae or core-collapse hypernovae. (4) Very massive
objects (i.e., M $>$ 100\,M$_{\odot}$) explode as pair-instability
hypernovae.  M$_{sw}$, the mass 
of interstellar gas collected by the explosion blast wave, depends 
on the explosion energy through Eq.~\ref{single_SN}.  m$_z$ is the mass of 
metals released by an exploding massive star with progenitor mass $m$.  
The metallicity [Fe/H] is derived assuming that the mass m$_z$ of 
metals is mixed with the mass M$_{sw}$ of collected interstellar gas}
\begin{center}
\begin{tabular}{ c c c c } \hline 
E$_{0}/10^{51}$\,ergs & 1 (SN) & 10 (HN) & 100 (HN)  \\ \hline
M$_{sw}$  & 5.1 $\times$ 10$^4$ & 5.1 $\times$ 10$^5$ & 5.1 $\times$ 10$^6$ \\ \hline \hline
(1)  (m,m$_z$)=(12,0.1)~  &   &   &    \\
{\rm [Fe/H]} $\simeq$ & $-$4 & -- & --  \\ \hline
(2)  (m,m$_z$)=(25,4)~~~ &  &   &    \\
{\rm [Fe/H]} $\simeq$ & $-$2.4 & $-$3.4 & $-$4.4  \\ \hline
(3)  (m,m$_z$)=(40,10)~~~ & &   &        \\ 
{\rm [Fe/H]} $\simeq$  & $-$2.1 & $-$3.1 & $-$4.1  \\ \hline
(4)  (m,m$_z$)=(200,100) & &   &        \\ 
{\rm [Fe/H]} $\simeq$  & -- & $-$2.0 & $-$3.0  \\ \hline
\end{tabular}
\end{center}
\end{table}   

The most significant feature of hypernova nucleosynthesis is their iron
production, this one being larger than in SNeII by a factor 2 to 10 (Nakamura 
et al.~2001).  This leads to small abundance ratios of $\alpha$ elements over 
iron.  Nakamura et al. (2001) thus suggested 
that the gas out of which the very metal-poor binary CS 22873-139 
([Fe/H] = $-$3.4) formed had been contaminated by the ejecta of an hypernova 
as this halo star shows almost solar [Mg/Fe] and [Ca/Fe] ratios.  On the other 
hand, while stars with $-2.5 \lesssim {\rm [Fe/H]} \lesssim 0$ show 
[Zn/Fe] $\simeq$ 0 (e.g., Primas et al.~2000), this abundance ratio is 
steadily increasing towards [Zn/Fe] $\sim$ 0.5 as the metallicity decreases,
for stars more metal-poor than [Fe/H] $\sim -2.5$ (e.g., Cayrel et al.~2003).
Umeda \& Nomoto (2002) notice that such a large [Zn/Fe] ratio arises 
naturally in their own hypernova model and thus conclude that 
core-collapse hypernovae are likely to have contributed to the early
Galactic chemical evolution.    \\

{\it Pair instability hypernovae.  }  
These very massive ($\simeq$ 140-260\,M$_{\odot}$) objects exploding with 
$E \gtrsim 10^{52}$\,ergs, they are also called hypernovae.  
Their name (i.e., ``pair instability'') refers to the
electron-positron pair instability process encountered during the central 
oxygen-burning stages (see Umeda \& Nomoto 2002, their appendix for a 
detailed description).

The main feature of these very massive 
stars is their ability to ``pass carbon and oxygen from the helium-burning 
core through the hydrogen-burning shell, in such a way that it is CNO 
processed to nitrogen before entering the hydrogen envelope''
(Carr, Bond \& Arnett 1984).  These stars thus produce 
large supersolar values of N/Fe, an effect not predicted by  
models of Galactic chemical enrichment based on stars less massive than 
100\,M$_{\odot}$.  Following their discovery of CS 22949-037, a
very metal-poor star ([Fe/H] = $-$3.8) showing
extreme nitrogen enhancement ([N/Fe] = 2.3),
Norris et al. (2002) suggested that this star may have been formed out
of gas polluted and compressed by such a very massive hypernova.  

On the other hand, in marked contrast with core-collapse hypernova yields,
Umeda \& Nomoto (2002) quote that pair-instability hypernovae are unlikely to
produce [Zn/Fe] ratios as large as in
very metal-poor stars where [Zn/Fe] ranges from solar up to $\simeq$0.5 
(Cayrel et al.~2003).  Actually,  the abundance ratio derived from 
their hypernova yields is of order [Zn/Fe]=$-$1.5, that is, 1 to 
2 dex smaller than in very metal-poor stars.
Additionally, Heger \& Woosley (2002) note the absence of the r-process 
in these massive objects, which conflicts with observations showing 
appreciable amounts of r-process elements in very metal-poor stars (e.g.,
Burris et al.~2000).  Based on these two arguments, i.e., the low abundance 
ratio [Zn/Fe] as well as the absence of r-process elements
in pair instability hypernova ejecta, Umeda \& Nomoto (2002) and 
Heger \& Woosley (2002) conclude that the abundances of very metal-poor 
stars cannot be ascribed to pair instability hypernovae only and
must include the contribution of an additional nucleosynthetic
component, namely those lower mass stars that make ``regular'' supernovae. 
We also note that pair-instability hypernovae release fairly large amounts of 
metals in the interstellar medium, of order 100~M$_{\odot}$ (Umeda 
\& Nomoto 2002, their Tables 15-18).  In fact, these stars disrupt completely 
when exploding, leaving no compact remnant after the explosion (i.e., 
no issue of ``mass-cut'' or ``fall-back'', see Heger \& Woosley 2002). 
The mixing of so large an amount of metals with a mass of primordial gas of 
$\simeq 5 \times 10^5 {\rm M}_{\odot}$ (E$_{0} = 10 \times 10^{51}$\,ergs) or 
$\simeq 5 \times 10^6 {\rm M}_{\odot}$ (E$_{0} = 100 \times 10^{51}$\,ergs) 
will lead to metallicities
of order [Fe/H] $\simeq -2$ or [Fe/H] $\simeq -3$, respectively.
Therefore, the metallicity of the most metal-poor stars in the Galactic halo
cannot be explained by pair-instability hypernovae. \\

\subsubsection{Single short-acting source of energy and triggered star formation}
The issue of whether isolated exploding massive or very massive stars can stimulate
the formation of new stars in the layers of interstellar gas they have swept
is not explicitely addressed in the papers mentioned in Sect.~3.2.1 and 3.2.2 
(e.g., Shigeyama \& Tsujimoto 1998, Argast et al.~2000, Norris et al.~2002).  
It is merely assumed that such stimulated star formation actually 
took place during the early Galactic stages.  

The model presented in this paper (see sect.~2) is not the most appropriate 
to address this issue quantitatively as it deals with a continuous input 
energy and not with a single short-acting source of energy.  Actually, 
supershells created in connection to single explosion and those created 
around a cluster of massive stars do not show the same expansion rate with 
time (see below).  We have seen in Sect.~\ref{sec:stimSF} how the radius and 
the velocity of the shell affect  its ability
to collapse through the shell surface density and the stretching of the forming
perturbation, respectively.  Hence, to solve the issue of whether a single 
supernova/hypernova triggers the formation of new stars requires to derive
a new temporal evolution of the shell radius, in the PGCC and in the protogalactic 
background, adequate to single explosion.  This is beyond the scope of the 
present paper whose main goal is to address the formation of stars with 
$-2.5 \lesssim {\rm [Fe/H]} \lesssim -1$ (i.e., the metallicity range of halo
GCs).  Some qualitative inferences can however be made. 
Efremov, Ehlerova \& Palous (1999) compared the propagation in an 
homogeneous medium of both kinds of supershells.  In case of an abrupt energy 
input, the early expansion of the supershell proceeds at a higher velocity than 
in the case of a continuous input of the same energy amount.  
Accordingly, in the former case, the shell of swept gas is more strongly 
stretched and this hampers the formation of an initial perturbation.  After
a few millions years however, the shell created in connection to an abrupt 
energy input
slows down and shows both a radius and a velocity lower than in case of 
a continuous energy input.  In contrast to the initial shell expansion, this
second stage favours the transverse collapse through a larger shell
surface density and a weaker stretching of the shell perturbed regions.
Whether the collapse will occur is actually not certain.  The source of the
energy input being point-like (i.e., its size is much smaller than the one of 
a cluster of massive stars), the initial amplitude of the transverse motions 
within the shell will be lowered accordingly, thus reducing the collapse. 

Comparing SNeII and hypernovae, hypernovae sweep a larger amount of gas thanks 
to their larger energy input (Eq.~\ref{single_SN}).  However, this also leads
to a larger shell radius.  Both effects acting on the shell surface density 
in opposite ways, only detail computations will tell us which class of objects
is the most efficient in forming new stars.  A $E_0 \sim 2 \times 
10^{52}$\,ergs hypernova being able to collect amount of gas as large as 
10$^6$\,M$_{\odot}$ (Eq.~$\ref{single_SN}$), one might think that 
pair-instability hypernovae could be related to the formation of GCs, at least 
the most metal-poor ones as such hypernovae can chemically enrich the 
initially pristine gas up to [Fe/H] $\simeq$ -2 (see Table 1).  
However, the [Zn/Fe] abundance ratio predicted for hypernova ejecta 
(i.e., [Zn/Fe] $<-1.5$, Umeda \& Nomoto 2002) shows a sharp discrepancy 
with the roughly solar value observed in halo field stars with [Fe/H] $\simeq -2$ 
(Primas et al.~2000) as well as in NGC6397, a halo GC with the same metallicity 
(Thevenin et al.~2001).  Therefore, the abundance patterns of stars in this
metallicity range cannot be ascribed to pair-instability hypernovae.

Obviously, it is not straightforward to conclude whether single massive star
explosions are able to trigger the formation of new stars.  More computations are 
required to answer a question which is certainly worthy of further investigation
owing to its potential link with very metal-poor stars.  
  
\subsubsection{External pollution onto PopIII stars}
Shigeyama, Tsujimoto \& Yoshii (2003) have recently suggested that some very 
metal-deficient stars may have been born as metal-free (PopIII) stars 
whose external layers were afterwards polluted by the accretion of chemically 
enriched interstellar gas.
The metallicity achieved by the external layers will depend on the 
metallicity and mass of the accreted gas, as well as on the stellar 
mass fraction with which the accreted gas has been mixed.  As a consequence, 
the chemical enrichment of stellar superficial layers will show up more
markedly in main sequence stars than in red giants, the 
convective envelope (i.e. the mixing zone) being almost two orders 
of magnitude larger in halo giants (about half the star mass) than in 
non-evolved metal-poor stars (about one per cent of the star mass).  Shigeyama
et al.~(2003) have shown that the extremely low metallicity 
([Fe/H]$\simeq -$5.3) of the giant HE0107-5240 (Christlieb et al.~2002) 
may originate from such a mechanism, that is, 
an external pollution while the star was on the main sequence followed 
by the dilution of the accreted material as the star started ascending the 
red giant branch.  Owing to their thin convective envelopes, polluted 
Pop III stars still on the main sequence would exhibit larger 
metallicities, i.e. of order [Fe/H]$\simeq -$3, although still 
lower than the most metal-poor GCs.   \\

\subsubsection{Star formation in dwarf galaxies}
Within the frame of a hierarchical model for halo
formation, Cote et al.~(1999) proposed that the excess of very metal-poor 
field stars with respect to GCs was formed in the most 
metal-deficient dwarf galaxies trapped within the Galactic potential 
well.  These galaxies are not expected to contribute to the GC population.  
Indeed, due to the luminosity-metallicity correlation observed among 
dwarf galaxies (Gilmore 2000, Mateo 2000), the most metal-poor dwarfs 
are also expected to be the least luminous.  Dwarf galaxies exhibit a 
luminosity threshold for hosting a GC system.  For instance, the Fornax
dwarf spheroidal galaxy is the faintest ($M_v \simeq -12.3$, Harris 1991)
galaxy known to harbour its own GC system.  Therefore, the dimmest dwarf 
galaxies, which are also the most metal-poor, may have contributed to 
the halo field and not to the halo GC system.  

\section{Caveats and Future work}
It is worthy of a mention that the model results presented above 
constitute a first and preliminary step in our computations of 
globular cluster 
star formation.  In fact, several aspects are still to be worked out.
We now give some examples of these, which our future developments will 
have to encompass.  
\begin{itemize}
\item[- ] As the shell collapse is followed until the perturbed and 
unperturbed surface densities reach similar values, a non-linear model 
would better describe the growth with time of the shell fragments.
In this respect, we note that Wunsch \& Palous (2001)
studied the transverse collapse of supershells propagating in the 
Galactic disc using both linear and non-linear models.  The comparison
of the results obtained in both cases, and for instance the time at 
which the shell gets fragmented, does not show significantly different results.
\item[- ] At some moments of its propagation through the hot protogalactic 
background, the shell undergoes transient accelerations (see 
Figs.~\ref{fig:RsHPB} and \ref{fig:RsHPBt1/3}) which may lead
to Rayleigh-Taylor instabilities.  Rayleigh-Taylor instabilities 
take place when a cold, dense gas (i.e. the supershell) is accelerated 
by pressure from a hot more rarefied gas (i.e. the bubble).
Spikes and globules of the colder fluid will tend to penetrate the warmer 
fluid (Spitzer 1978).  While this effect may disrupt the 
shell and, thus, prevent the process of triggered star formation, 
it can also lead to local overdensities in the shell material, accelerating 
thereby star formation.  It will therefore be of interest to estimate
whether the shell transient accelerations may trigger Rayleigh-Taylor
instabilities, how far these ones develop and how the process of stimulated 
star formation within the shell is affected.   
\item[- ] The computation of the shell expansion law can still be improved 
as it does not yet include energy losses, such as those induced by viscous 
processes.  For given number of SNeII and external pressure, these effects 
may lower the radius of the shell with respect to what has been computed 
in Sect.~2.2 and, thus, increase the shell surface density and the 
probability of star formation.  If the formation of stars may be 
triggered even at pressures lower than $P_h = 10^{-11}$\,dyne.cm$^{-2}$, 
metallicities lower than [Fe/H]$\simeq -$2.8  may be achieved.  
\end{itemize} 

\section{Summary}
This paper has presented the results of simulations dedicated to the
transverse collapse of shells of gas resulting from the sweeping of gaseous 
GC progenitors by SNeII.  In these simulations, the growth of an initial
perturbation in the shell surface density is followed by solving the linear perturbed
equations of continuity and motion for transverse flows in a spherical shell
of gas.
Such a collapse depends on several parameters, namely the number of SNeII,
the background presure, the sound speed of the shell gas and the initial 
conditions of the perturbation (i.e. the number of clumps, the initial
perturbed surface density and velocity and the corresponding phase
difference).  All these parameters have been discussed in turn.  The
results show that the pressure $P_h$ of the hot protogalactic background 
($P_h \sim 10^{-10}$\,dyne.cm$^{-2}$, Fall \& Rees 1985,
Murray \& Lin 1992) and the numbers $N$ of SNeII allowed by the disruption
criterion (i.e. smaller than 200, Paper I) can indeed lead to a successful 
shell transverse collapse, and thereby to the formation of new stars,
assuming some reasonable initial conditions for the perturbation
(see Fig.~8).  The metallicities achieved in the shells able to collapse 
agrees with the metallicity range of Galactic halo GCs, namely, 
$-2.5 \lesssim {\rm [Fe/H]} \lesssim -1$.
Furthermore, while $N$ and $P_h$ determine the metallicity achieved through
self-enrichment, they also control the probability of triggered star formation 
and the ability of these second generation stars to form a bound GC.  Such
a property is the most interesting since it opens the way to the 
understanding of the halo metallicity distribution functions, for both stars 
and clusters.

\bsp

\section*{Acknowledgments}
I am grateful to Richard Scuflaire for fruitful advices and discussions.
Supports from P\^ole d'Attraction Interuniversitaire through grant P5/36 
(SSTC, Belgium) and from the European Commission through grant 
HPMT-CT-2000-00132 are gratefully acknowledged.

\label{lastpage}


\begin{thebibliography}{99}
\bibitem{abel}
Abel T., Bryan G.L., Norman M.L., 2002, Science, 295, 93
\bibitem{argast}
Argast D., Samland M., Gerhard O.E., Thielemann F.-K., 2000, A\&A, 356, 873
\bibitem{bromm}
Bromm V., Coppi P.S., Larson R.B., 1999, ApJL, 527, 5 
\bibitem{brown1}
Brown J.H., Burkert A., Truran J.W. 1991, ApJ 376, 115 
\bibitem{brow2}
Brown J.H., Burkert A., Truran J.W. 1995, ApJ 440, 666
\bibitem{burris}
Burris D.L., Pilachowski C.A., Armandroff T.E., Sneden C., Cowan J.J., Roe H., 2000,
ApJ, 544, 302
\bibitem{campbell}
Campbell B., Hunter D.A., Holtzman J.A., Lauer T.R., Shayer E.J., 
Code A., Faber S.M., Groth E.J., Light R.M., Lynds R., O'Neil E. Jr.,
Westphal J.A. 1992, AJ 104, 1721
\bibitem{carr}
Carr B.J., Bond J.R. \& Arnett W.D. 1984, ApJ 277, 445
\bibitem{castor}
Castor J., McCray R., Weaver R., 1975, ApJ 200, L107 
\bibitem{cayrel}          
Cayrel R. 1986, A\&A 168, 81 
\bibitem{cayrel03}
Cayrel R., Depagne E., Spite M., Hill V., Spite F., Francois P., Plez B., Beers T., 
Primas F., Andersen J., Barbuy B., Bonifacio P., Molaro P., Nordstrom B., 2003,
A\&A, accepted for publication
\bibitem{christlieb02}
Christlieb N., Bessell M.S., Beers T.C., Gustafsson B., Korn A., 
Barklem P.S., Karlsson T., Mizuno-Wiedner M., Rossi S. 2002, Nature 419, 904
\bibitem{christlieb03}
Christlieb N., Gustafsson B., Korn A., Barklem P.S., Beers T.C., Bessell M.S., 
Karlsson T., Mizuno-Wiedner M., 2003, ApJ, accepted for publication
\bibitem{comeron}
Comeron F., Torra J. 1994, ApJ 423, 652
\bibitem{cote}
Cote P. 1999, AJ 118, 406
\bibitem{dinge}
Dinge D., 1997, ApJ, 479, 792
\bibitem{efremova}
Efremov Y.N., Elmegreen B.G. 1998, MNRAS 299, 643
\bibitem{efremovb}
Efremov Yu.N., Ehlerova S., Palous J. 1999, A\&A 350, 475
\bibitem{ehlerova}
Ehlerova S. \& Palous J. 2002, MNRAS 330, 1022
\bibitem{elmegreen}
Elmegreen B.G. 1994, ApJ 427, 384
\bibitem{fallrees}
Fall S.M., Rees M.J. 1985, ApJ 298, 18
\bibitem{galama}
Galama T., et al. 1998, Nature 395, 670
\bibitem{gilmore}
Gilmore, G. 2000, in ASP Conf. Ser. 230, Galaxy Disks and Disk Galaxies, 
eds. J.G. Funes \& E.M. Corsini, 3
\bibitem{gnedin}
Gnedin O.Y., Ostriker J.P., 1997, ApJ, 474, 223
\bibitem{harris}
Harris, W.E. 1991, ARA\&A, 29, 543
\bibitem{hp}
Harris W.E., Pudritz R.E. 1994, ApJ 429, 177 
\bibitem{heger}
Heger A. \& Woosley S.E. 2002, ApJ 567, 532
\bibitem{karlsson}
Karlsson T. \& Gustafsson B. 2001, A\&A 379, 461
\bibitem{kennicutt}
Kennicutt R.C. Jr. 1989, ApJ 344, 685
\bibitem{lada}
Lada E.A., Strom K.M., Myers P.C., 1993, in: Protostars and planets III, p245
\bibitem{laird}
Laird, J.B., Carney B.W. \& Latham D.W. 1988, ApJ 95, 1843 
\bibitem{larson}
Larson R.B. 1988, in: The Harlow-Shapley Symposium on Globular Cluster Systems 
in Galaxies, eds. Grindley J.E. \& Davis Philip A.G., Dordrecht, Kluwer Academic 
Publishers, p311
\bibitem{mateo}
Mateo, M. 2000, in The First Stars (Proceedings of the MPA/ESO Workshop 
held at Garching, Germany, 4-6 August 1999), eds. A. Weiss, T.G. Abel \& 
V. Hill, Springer, 283
\bibitem{mccray}
McCray R., Kafatos M. 1987, ApJ 317, 190
\bibitem{murray}
Murray S.D., Lin D.N.C. 1992, ApJ 400, 265  
\bibitem{nakamura99}
Nakamura F., Umemura M., 1999, ApJ, 515, 239 
\bibitem{nakaT01}
Nakamura T., Umeda H., Iwamoto K., Nomoto K., Hashimoto M., Hix W.R.,
Thielemann F.-K. 2001, ApJ 555, 880
\bibitem{nakamura01}
Nakamura F., Umemura M., 2001, ApJ, 548, 19 
\bibitem{norris00}
Norris J.E., Beers T.C. and Ryan S.G. 2000, ApJ 540, 456
\bibitem{norris02}
Norris J.E., Ryan S.G., Beers T.C., Aoki W. \& Ando H. 2002, ApJL 569, 107
\bibitem{opik}
Opik E.J. 1953, IrAJ 2, 2190
\bibitem{palla}
Palla F., Salpeter E.E., Stahler S.W., 1983, ApJ, 271, 632
\bibitem{parm1}
Parmentier G., Jehin E., Magain P., Neuforge C., Noels A., Thoul A.A. 1999,
A\&A 352, 138 (Paper I)
\bibitem{parm2} 
Parmentier, G., Jehin, E., Magain, P., Noels, A., \&
Thoul, A. 2000, A\&A, 363, 526
\bibitem{parm3} 
Parmentier G. \& Gilmore G. 2001, A\&A 378, 97
\bibitem{primas}
Primas F., Brugamyer E., Sneden C., King J.R., Beers T.C., Boesgaard A.M.
\& Deliyannis C.P. 2000, in: The First Stars, eds. Weiss A., et al., Springer, p51
\bibitem{rosenberg}
Rosenberg A., Saviane I., Piotto G., Aparicio A., 1999, AJ, 118, 2306
\bibitem{rubio}
Rubio M., Barba R.H., Walborn N.R., Probst R.G., Garcia J., Roth M.R.
1998, AJ 116, 1708
\bibitem{shigeyama}
Shigeyama T.., Tsujimoto T. \& Yoshii, Y. 2003, ApJL 586, 57
\bibitem{smith}
Smith G.H., 1999, ApJL, 526, 21
\bibitem{spitzer}
Spitzer L. 1978, Physical Processes in the Interstellar Medium, 
Wiley-InterScience
\bibitem{thevenin}
Thevenin F., Charbonnel C., de Freitas Pacheco J.A., Idiart T.P., Jasniewicz G.,
de Laverny P. and Plez B. 2001, A\&A 373, 905
\bibitem{tsujimoto95}
Tsujimoto T., Nomoto, K., Yoshii Y., Hashimoto M., Yanagida S. and 
Thielemann F.-K. 1995, MNRAS 277, 945
\bibitem{tsujimoto98}
Shigeyama T. \& Tsujimoto T. 1998, ApJL 507, 135 
\bibitem{tsujimoto_b}
Tsujimoto T, Shigeyama T., Yoshii Y. 1999, ApJL 519, 63 
\bibitem{umeda}
Umeda H. \& Nomoto K. 2002, ApJ 565, 385
\bibitem{vietri}
Vietri M., Pesce E., 1995, ApJ, 442, 618
\bibitem{wasserburg}
Wasserburg G.J. \& Qian Y.-Z. 2000, ApJL 531, 33
\bibitem{walter}
Walter F., Kerp J., Neb D., Brinks E., Klein U., 1998, ApJ, 502, L143 
\bibitem{weaver}
Weaver R., McCray, R., Castor, J., 1977, ApJ 218, 377 
\bibitem{woosley}
Woosley S.E. \& Weaver T.A. 1995, ApJS 101, 181
\bibitem{wunsch}
Wunsch R. \& Palous J. 2001, A\&A 374, 746
\bibitem{zinn}
Zinn R. 1985, ApJ 293, 424
\end{thebibliography}
\end{document}